\documentclass{aa}

\usepackage{amsmath}
\usepackage{amssymb}
\usepackage[usenames, dvipsnames]{color}
\usepackage{footmisc}
\usepackage[modulo,switch]{lineno}
\usepackage{graphicx}
\usepackage{multirow}
\usepackage{natbib}
\usepackage{nicefrac}
\usepackage{pdfpages}
\usepackage[rightcaption]{sidecap}
\usepackage{txfonts}
\usepackage{units}
\usepackage{url}

\usepackage{hyperref}
\hypersetup{
    final=true,
    pageanchor=true,
    colorlinks=true,
    breaklinks=true,
    linkcolor=blue,
    citecolor=blue,
    urlcolor=blue,
    pdfpagemode=UseNone,
    pdftitle={Diercke2019},
    pdfauthor={Diercke},
    pdfsubject={Solar Physics},
    pdfkeywords={}}

\bibpunct{(}{)}{;}{a}{}{,} 
\newcommand{\arcdeg}{\mbox{$^{\circ}$}}
\newcommand\arcsecdot{\mbox{$^{\prime\prime}$}\hspace{-0.15cm}.\,} 

\begin{document} 

\title{Dynamics and connectivity of an extended arch filament system}

\author{%
    A.\ Diercke\inst{1,2} \and 
    C.\ Kuckein \inst{1} \and
    C.\ Denker \inst{1}}
\institute{%
    Leibniz-Institut f\"ur Astrophysik Potsdam,
    An der Sternwarte 16,
    14482 Potsdam, Germany\\
    \email{adiercke@aip.de}
\and
    Universit\"at Potsdam,
    Institut f\"ur Physik und Astronomie,
    Karl-Liebknecht-Stra\ss{}e 24/25,
    14476 Potsdam,
    Germany}

\date{Version \today; Received March 29, 2019; accepted July 30, 2019}

 
\abstract
{}
{In this study, we analyzed a filament system, which expanded between moving magnetic features (MMFs) of a decaying sunspot and opposite flux outside of the active region from the nearby quiet-Sun network. This configuration deviated from a classical arch filament system (AFS), which typically connects two pores in an emerging flux region. Thus, we called this system an extended AFS. We contrasted classical and extended AFSs with an emphasis on the complex magnetic structure of the latter. Furthermore, we examined the physical properties of the extended AFS and described its dynamics and connectivity.}
{The extended AFS was observed with two instruments at the Dunn Solar Telescope (DST). The Rapid Oscillations in the Solar Atmosphere (ROSA) imager provided images in three different wavelength regions, which covered the dynamics of the extended AFS at different atmospheric heights. The Interferometric Bidimensional Spectropolarimeter (IBIS) provided spectroscopic H$\alpha$ data and spectropolarimetric data that was obtained in the near-infrared (NIR) \mbox{Ca\,\textsc{ii}} $\lambda$8542\,\AA\ line. We derived the corresponding line-of-sight (LOS) velocities and used \mbox{He\,\textsc{ii}} $\lambda$304\,\AA\ extreme ultraviolet (EUV) images of the Atmospheric Imaging Assembly (AIA) and LOS magnetograms of the Helioseismic and Magnetic Imager (HMI) on board the Solar Dynamics Observatory (SDO) as context data.}
{The NIR \mbox{Ca\,\textsc{ii}} Stokes-$V$ maps are not suitable to definitively define a clear polarity inversion line and to classify this chromospheric structure. Nevertheless, this unusual AFS connects the MMFs of a decaying sunspot with the network field. At the southern footpoint, we measured that the flux decreases over time. We find strong downflow velocities at the footpoints of the extended AFS, which increase in a time period of 30\,minutes. The velocities are asymmetric at both footpoints with higher velocities at the southern footpoint. An EUV brigthening appears in one of the arch filaments, which migrates from the northern footpoint toward the southern one. This activation likely influences the increasing redshift at the southern footpoint.}
{The extended AFS exhibits a similar morphology as classical AFSs, for example, threaded filaments of comparable length and width. Major differences concern the connection from MMFs around the sunspot with the flux of the neighboring quiet-Sun network, converging footpoint motions, and longer lifetimes of individual arch filaments of about one hour, while the extended AFS is still very dynamic.}


\maketitle


\section{Introduction}

Filaments are clouds of dense plasma, which reside at chromospheric or coronal heights in the solar atmosphere. Due to differences in temperature and density between the filament and the coronal plasma and absorption of the underlying radiation, filaments materialize as dark structures on the solar disk. Observed above the limb, filaments appear in emission as loops and are called prominences \citep[e.g., ][]{Martin1998a,Mackay2010}. They are formed above the polarity inversion line \citep[PIL,][]{Mackay2010}, which is defined as the boundary between positive and negative polarities of the line-of-sight (LOS) magnetic field. Filaments, which are embedded in active regions, often close to sunspots, are called active region filaments \citep{Tandberg2001}. They are usually shorter than their counterparts in the quiet-Sun and reach lengths between 10--100\,Mm but the plasma contained in the filament can reach higher velocities of 30\,km\,s$^{-1}$ along the spine of the filament. Compared to quiet-Sun filaments, the magnetic field strength in active region filaments is higher with values of 50--200\,G. However, even field strengths of up to 600--700\,G have been measured \citep[e.g.,][]{Kuckein2009}. The lifetime of active region filaments is in the range of a few hours to days \citep{Mackay2010}. In contrast, single threads have a length of about 11\,Mm, a width of 1--2\,Mm, and lifetimes in the range of 10--20\,minutes \citep{Tandberg2001}.

\begin{figure*}
\center
\includegraphics[width=1\textwidth]{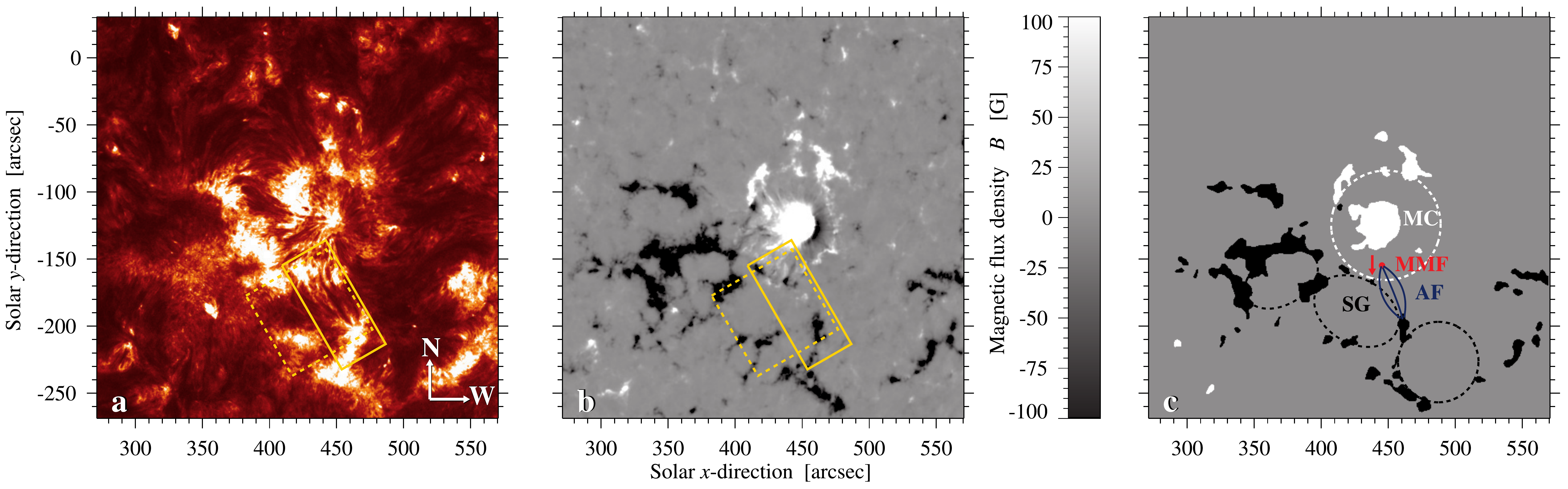}
\caption{Overview of observed target on 2013~January~20. (a) Detailed view of active region NOAA 11658 at 17:35:55~UT observed in the EUV at $\lambda$304\,\AA. The sunspot is located in the middle of the FOV. The solid yellow box indicates the FOV of IBIS and the dashed yellow box indicates the FOV of ROSA. Both contain the targeted extended AFS. (b) Averaged magnetogram between 14:00~UT and 22:00~UT. (c) Schematic overview of NOAA 11658 with the moat cell (MC) of the moat flow around the sunspot, an MMF in the moat cell, and arch filaments (AF) along a supergranule (SG). The arrow indicates the propagation direction of the MMF.\label{FIG:SolMon}}
\end{figure*}

Unlike filaments, arch filament systems \citep[AFSs,][]{Bruzek1967, Bruzek1969} bridge the PIL while connecting two opposite magnetic polarities. Classical AFSs are observed within emerging flux regions \citep[EFRs,][]{Zirin1972} and young bipolar spot groups; they are thus associated with the evolution of bipolar spot groups \citep{Bruzek1967, Frazier1972, Ma2015}. \citet{GonzalezManrique2017} describe an AFS connecting emerging flux of micro-pores with quiet-Sun magnetic fields of opposite polarity. In their study, the small-scale filaments indicate that new flux emerges in the form of cool loops. The footpoints are rooted in bright plage regions \citep{Bruzek1967}. The AFSs are very dynamic, and LOS velocity maps show strong downflows in the chromosphere at the footpoints of the AFS and upflows at the loop tops. Average velocities are in the range 20--50\,km\,s$^{-1}$ \citep{Bruzek1967, GonzalezManrique2018}, and the upflows are in the range 1.5--20\,km\,s$^{-1}$ at the loop tops \citep{GonzalezManrique2017, GonzalezManrique2018, Balthasar2016}. Typical AFSs contain bundles of 5 to 15 individual small filamentary structures with a width of only a few megameters \citep{Bruzek1967}. The common width and length of the entire system is about 20\,Mm and 30\,Mm, respectively. The typical lifetime of the individual structures is approximately 30\,minutes, while the whole system can have a lifetime of a few days \citep{Bruzek1967, GonzalezManrique2017b, GonzalezManrique2018}. Significant changes in the system can appear within several hours. In any case, the disappearance of AFSs is not always related to changes in the accompanying sunspots.

Arch filament systems are chromospheric features that are best observed in H$\alpha$ filtergrams. Here, the dark structures connect to regions of bright plage \citep{Ma2015}. In addition, the filament can be observed in different calcium lines, such as the NIR \mbox{Ca\,\textsc{ii}} $\lambda$8542\,\AA\ line, although the AFS appears less prominent. \citet{Bruzek1969} reports that the size of an AFS and the cells of the \mbox{Ca\,\textsc{ii}\,K} network have approximately the same size. They concluded that an AFS may cover one supergranule.

In high-resolution images, bright points become sometimes visible at the boundary of granules, for example, in the Frauenhofer G-band at around $\lambda$4307\,\AA\ \citep{Muller2001}. In active regions, bright points surround individual or ensembles of granules with a diameter of 2--3\arcsec. These bright points are strongly related to magnetic flux concentrations and are thought to be the footpoints of magnetic loops, which can extend up into the corona \citep{Rimmele2004}. Bright points are typically observed around AFSs,  especially close to the footpoints of the filaments \citep{Bruzek1967}.

Moving magnetic features \citep[MMFs,][]{Harvey1973b} are small-scale phenomena, which appear in the moat of a sunspot \citep{Stix2004}. \citet{Shine2001} distinguished between three different types. Type~I denotes bipolar features, which often appear in continuation of dark penumbral filaments \citep{MartinezPillet2002}. Type~II refers to unipolar features with the same polarity as the sunspot, which are often associated with the decay of the sunspot. Type~III relates to unipolar features but with opposite polarity as the sunspot. They are also connected to penumbral filaments, similar to type~I MMFs.

\citet{Zirin1974} stated that a classical AFS, which forms above an EFR, has no connection to the surrounding network. We present the analysis of an atypical AFS, which connects the flux of MMFs in the vicinity of a sunspot with the flux of opposite polarity in the network magnetic field. Therefore, we compare properties, such as length, width, lifetime, and amount of flux, of this AFS using multiple instruments and wavelengths, with those of classical AFSs and active region filaments, searching for common properties. In Sect.~\ref{sec:obs}, we describe the ground-based observations with different instruments and in different wavelengths. In Sect.~\ref{sec:meth}, we provide the details of the data reduction. Finally, the results and their discussion are presented in Sects.~\ref{sec:results} and \ref{sect:disc}.

\begin{figure*}[t]
\includegraphics[width=1.\textwidth]{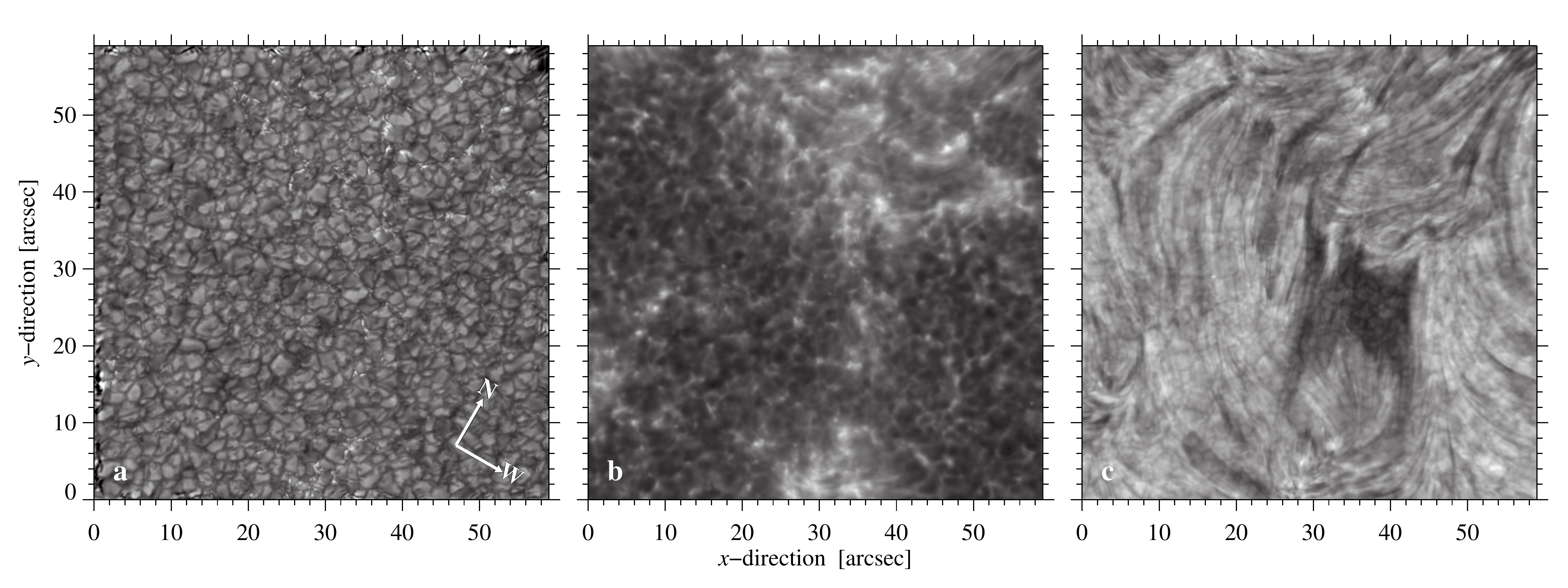}
\caption{Best ROSA images on 2013~January~20: (a) G-band image showing the
    photosphere at 17:21:39~UT, where some restoration artifacts are
    visible at the borders. (b) \mbox{Ca\,\textsc{ii}\,K} 
    image depicting the upper photosphere and lower chromosphere at 
    17:21:34~UT. (c) H$\beta$ image displaying the chromosphere at 
    17:21:22~UT. The arrows indicate the direction of solar north and west.\label{FIG:ROSA}}
\end{figure*}


\section{Observations}\label{sec:obs}

In this work, we used data from the Dunn Solar Telescope \citep[DST,][]{Bhatnagar2005}, which is located in Sunspot, New Mexico (USA) in the Sacramento Mountains. We analyzed data from two instruments, that is, the Rapid Oscillations in the Solar Atmosphere \citep[ROSA,][]{Jess2010} and the Interferometric Bidimensional Spectropolarimeter \citep[IBIS,][]{Cavallini2006, Reardon2008}. These two instruments provide in the optical wavelength range access to high-resolution images and to imaging spectropolarimetric data. The yellow squares in Fig.~\ref{FIG:SolMon} indicate the size and orientation of the observed field-of-view (FOV) for ROSA and IBIS.

The ROSA instrument was used to observe in three different wavelengths, that is, the Frauenhofer G-band $\lambda$4305.5\,\AA, and in two  strong chromospheric absorption lines H$\beta$ $\lambda$4861.3\,\AA\ and \mbox{Ca\,\textsc{ii}\,K} $\lambda$3933.7\,\AA. All three filters show different layers of the solar atmosphere. The G-band filter shows the photosphere, the H$\beta$ images unveils the chromosphere, similar to the H$\alpha$ observations, and the \mbox{Ca\,\textsc{ii}\,K} filtergrams cover the upper photosphere and lower chromosphere. The ROSA images were taken with CCD cameras, which have a detector with $1004 \times 1002$~pixels and a pixel size of $0\arcsecdot06\,\times\,0\arcsecdot06$. Because of the different transmission and bandpass of the filters, different exposure times were needed in the observations. The G-band, H$\beta$, and \mbox{Ca\,\textsc{ii}\,K} data had an exposure time of 17\,ms, 450\,ms, and 370\,ms, respectively. Ultimately, we obtained cadences of 11.8\,s, 29.4\,s, and 26.5\,s for reconstructed G-band, H$\beta$, and \mbox{Ca\,\textsc{ii}\,K} images, respectively.

The IBIS observing sequence scanned sequentially the H$\alpha$ $\lambda$6562.86\,\AA\ and \mbox{Ca\,\textsc{ii}}~$\lambda$8542.21\,\AA\ lines. The exposure time for each channel was 80\,ms. The wavelength steps are not equidistant and vary between 0.24\,\AA\ in the wings and 0.04\,\AA\ in the core. The \mbox{Ca\,\textsc{ii}}~$\lambda$8542\,\AA\ (hereafter NIR~\mbox{Ca\,\textsc{ii}}) spectropolarimetric observations contain different  polarization states yielding the four Stokes parameters. The H$\alpha$ data are only spectroscopic observations. The dual-beam mode of the spectropolarimeter reduces the effective  FOV to $35.3\arcsec \times 84.7\arcsec$.  The image scale is about \mbox{0\arcsecdot099}\,pixel$^{-1}$.

All data were obtained in a DST service-mode campaign on 2013~January~20, and the main target was active region NOAA 11658, which contained an axisymmetric sunspot that is often called a theoretician's spot. This sunspot belongs according to the McIntosh classification to class Hsx. These spots are usually the remaining leading spots of large bipolar regions and survive after the rest of the group vanished from the solar surface. They remain unchanged for several weeks \citep{Stix2004}, as did the spot in the current case. The bipolar character of the active region is still visible in magnetograms. The sunspot was surrounded by filamentary structures (yellow rectangles in Fig.~\ref{FIG:SolMon}). The central heliographic coordinates of the active region are 13\arcdeg\,S and 32\arcdeg\,W. The observed AFS was located at 15.2\arcdeg\,S and 27.6\arcdeg\,W at the beginning of observations at 15:48~UT and connected flux around the sunspot with flux from a quiet-Sun region close by. The AFS resided at the border of a supergranule. In Fig.~\ref{FIG:SolMon}c, we present a schematic overview of NOAA 11658.

Two instruments on board the Solar Dynamics Observatory \citep[SDO, ][]{Pesnell2012}, that is, the Atmospheric Imaging Assembly \citep[AIA, ][]{Lemen2012} and Helioseismic and Magnetic Imager \citep[HMI, ][]{Scherrer2012}, provide context data for the interpretation of the high-resolution DST data. The data was rescaled and derotated with standard SDO and image processing routines as described by \citet{Diercke2018}. In addition, the contrast in the AIA images was enhanced with noise adaptive fuzzy equalization \citep[NAFE,][]{Druckmueller2013}. They cover the time between 14\,--\,22\,UT on 2013~January~20. We present an overview of the active region in Fig.~\ref{FIG:SolMon} indicating the FOV of IBIS and ROSA. In the AIA\,$\lambda304$\,\AA\ image (hereafter AIA image), we see several AFSs around the main sunspot of the group (Fig.~\ref{FIG:SolMon}a). The averaged magnetogram shows the magnetic configuration of the active region and the surrounding network (Fig.~\ref{FIG:SolMon}b).


\section{Data reduction and analysis} \label{sec:meth}

\subsection{Data reduction of ROSA images}

First, we calibrate the data with the averaged dark and flat-field
frames. To evaluate the image quality of the time-series for later image 
reconstruction, we use the Median Filter-Gradient Similarity \citep[MFGS, 
][]{Deng2015} method with an implementation as described in \citet{Denker2018}, which 
is part of the sTools software package \citep{Kuckein2017IAU}. The best 
image quality for all three wavelengths occurs at around the same time, 
indicating that the seeing conditions were excellent at that time. The direct
comparison of the MFGS values at different wavelengths is difficult, because
the wavelength dependence is not known, yet. Other factors such as the amount of 
fine structure of the observed feature and the signal-to-noise ratio, influence 
the MFGS value as well. However, an assessment of image quality is possible 
within the individual time-series.

For image restoration, we rely on the MFGS values to identify the best images. We first select and align the 16 best images of each 30-second sequence as input for multi-frame blind deconvolution \citep[MFBD,][]{Loefdahl2002}. The MFBD  method uses blind deconvolution, in which the point spread function (PSF) is not known. The PSF is found iteratively by solving an optimization problem for a mosaic of isoplanatic patches \citep{Gonzalez2002, vanNoort2005, Delacruzrodriguez2015}. In the end, we obtain restored images with higher spatial resolution and contrast (Fig.~\ref{FIG:ROSA}), which are also corrected for stray light. Still some artifacts remain at the borders of the G-band images (Fig.~\ref{FIG:ROSA}a) because the images were wrapped at the borders to preserve the FOV in the restored images.

\begin{figure}[t]
\includegraphics[width=1\columnwidth]{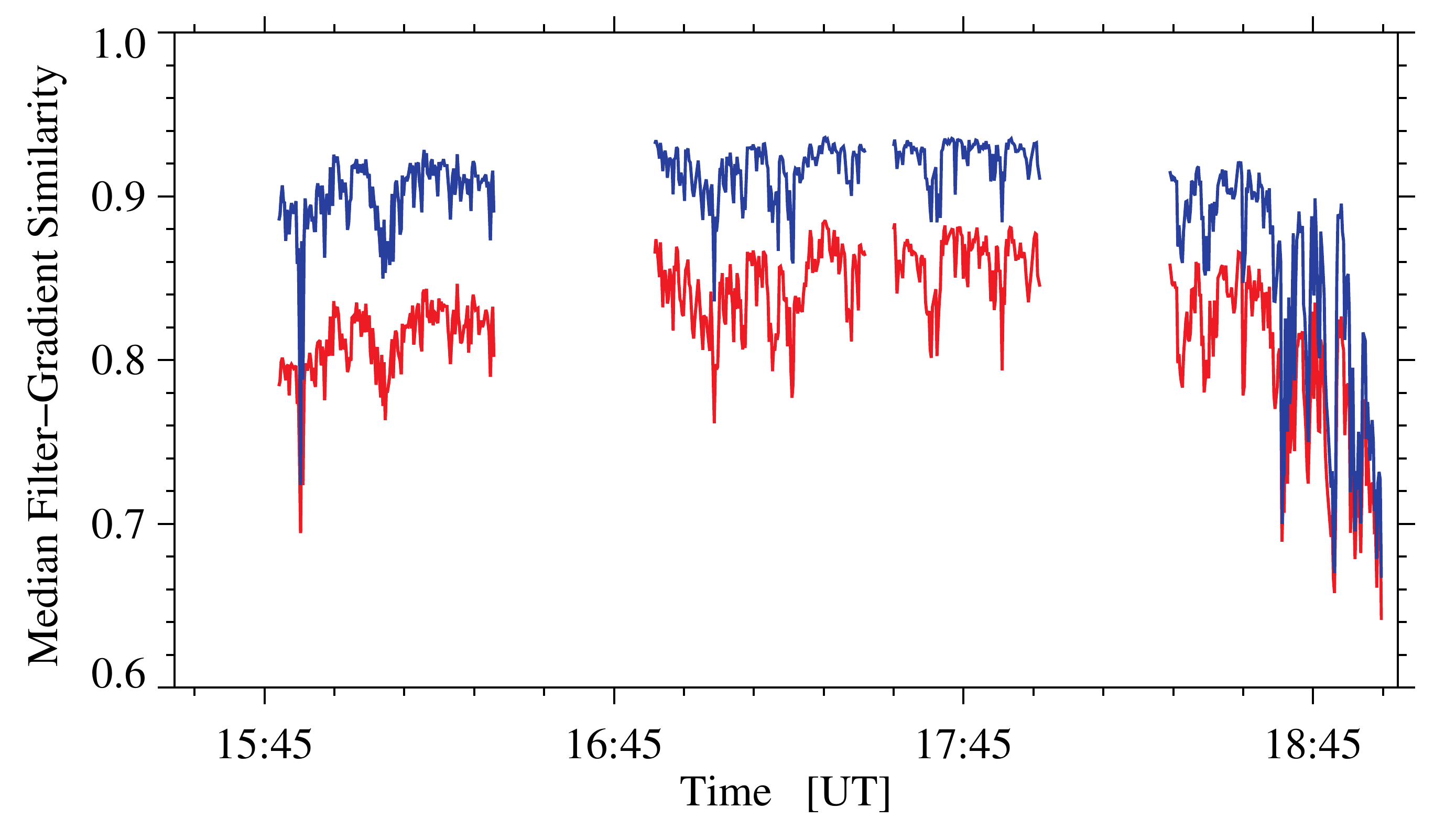}
\caption{Image quality of G-band images derived with the MFGS method for the
     entire observing period. The red curve refers to calibrated G-band images,
     whereas the blue profile corresponds to MFGS values after image restoration.
     At around 18:21~UT the target changed from the arch filament system to the
     nearby sunspot.\label{FIG:Quality}}
\end{figure}

A time-series analysis requires a more or less coherent time-series without any larger gaps. The best seeing conditions were encountered between 16:52~UT and 17:58~UT (Fig.~\ref{FIG:Quality}). We thus concentrate on these data covering the most stable time period with the best seeing. Unfortunately, this time-series still contains some smaller gaps. Finally, we rotate the images by 180\arcdeg\ and transpose them. The orientation of the images with respect to the standard disk-center coordinates is indicated by the arrows in Fig.~\ref{FIG:ROSA}a.

\subsection{Data reduction of IBIS spectropolarimetric observations}

The IBIS data reduction pipeline is described in \citet{Criscuoli2014}. The reduction is carried out for each filter separately, using different steps for spectroscopic and spectropolarimetric data of the narrow-band channel \citep{Judge2010}. Basic data processing includes dark and flat-field correction, alignment of narrow- and broad-band images, and creation of metadata (wavelength position, time of observations, light level, etc.). In the next step, the blueshift of the narrow-band data is corrected, which is introduced by the classical mount of the Fabry-P\'erot interferometers. Finally, image distortions are removed using a destretching algorithm. The spectropolarimetric NIR~\mbox{Ca\,\textsc{ii}} data requires three additional reduction steps to derive the four elements of the Stokes vectors. In the first step, the polarization calibration curve is generated. In the second step, the polarization curve is used to calculate the polarimeter response function. In the last step, this function is used to demodulate and calibrate the images for different polarization states and to derive the Stokes vectors. Ultimately, we obtain a narrow-band image for each of the four elements of the Stokes vector at each of the 25 wavelength positions. All images are cropped to  $384 \times 900$~pixels, which corresponds to a FOV of about $38\arcsec \times 89\arcsec$.

\subsection{Prefilter curve for spectroscopic IBIS data}\label{sect:prefilter}

\begin{figure}
\includegraphics[width=1\columnwidth]{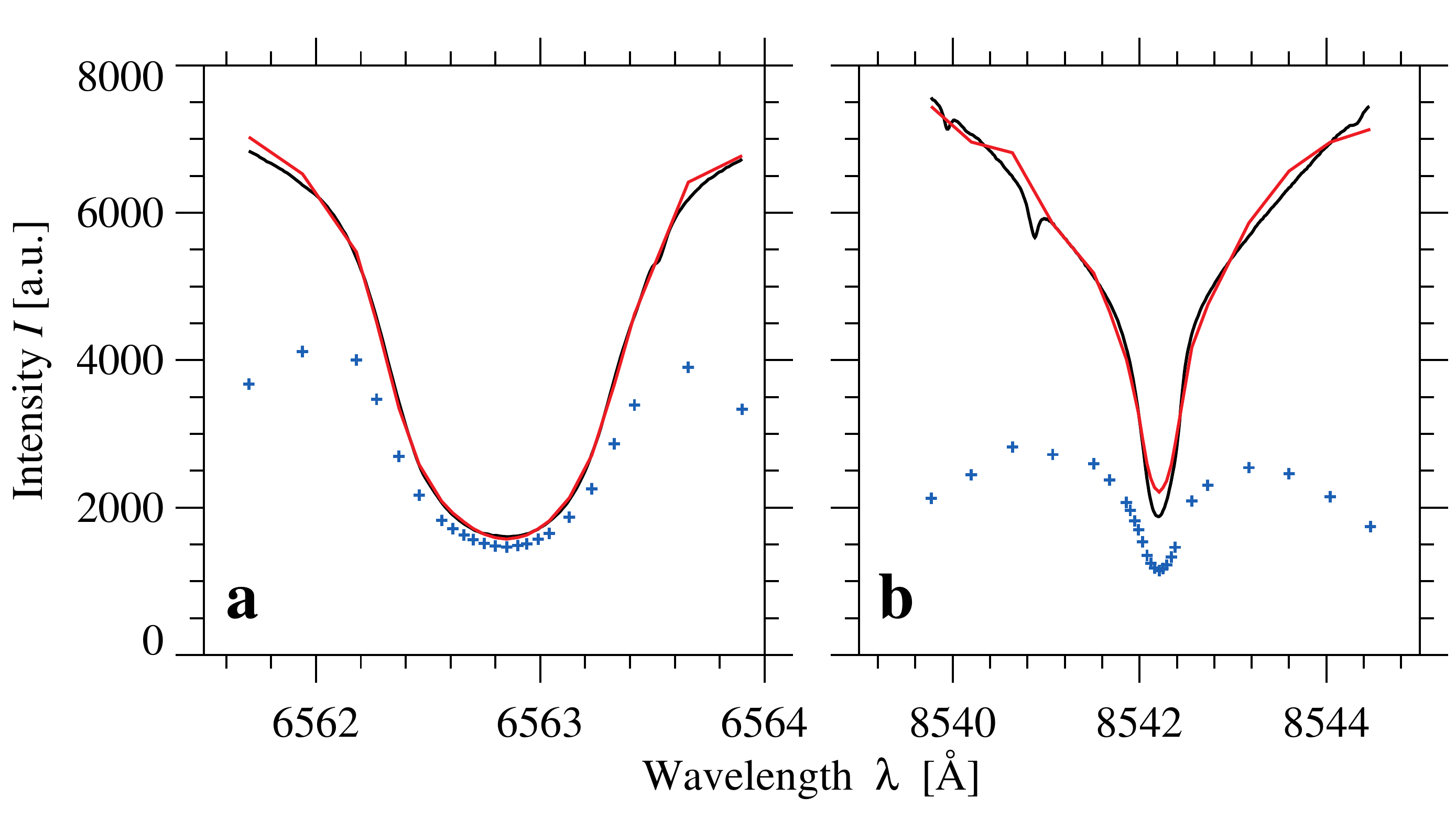}
\caption{FTS spectrum (black line), original IBIS spectrum (blue crosses), and corrected IBIS spectrum after division by the prefilter curve (red line) for (a) H$\alpha$ and (b) NIR~\mbox{Ca\,\textsc{ii}}.\label{FIG:prefilter}}
\end{figure}

Each filter has a specific transmission curve, which influences the spectral profile. We start with an average spectrum derived from the flat-field frames. The spectrum is not equidistantly sampled. Therefore, we compare the IBIS spectrum with a disk-integrated reference spectrum taken from the Kitt Peak Fourier Transform Spectral \citep[FTS,][]{Brault1985} atlas. The FTS spectrum is selected for the observed wavelength range of the IBIS spectrum, and the latter is interpolated to an equidistant grid matching the dispersion $d_\mathrm{FTS} = 0.02$\,\AA\ of the FTS atlas. Wavelength shifts between FTS and IBIS spectra are determined by parabola fits to the line cores, and they are corrected in the previous interpolation as well. To derive  the prefilter curve, we divide the two spectra and fit a Gaussian function to this ratio. Subsequently, all spectral profiles are divided, pixel by pixel, by the prefilter curve. Figure~\ref{FIG:prefilter} shows the uncorrected spectrum (blue), the FTS spectrum (black), and the corrected spectrum after division by the prefilter curve (red). 

The spectroscopic data contain several scans through the spectral line of H$\alpha$ in the range 6561.46--6564.14~\AA\ and of NIR~\mbox{Ca\,\textsc{ii}} in the range 8539.77--8544.47\,\AA. The steps between the scans are non-equidistant.  All spectral profiles in the data cube were divided by the prefilter curve. Furthermore, the spectrum was interpolated to an equidistant grid with a step size corresponding to the smallest interval of the grid, that is, 0.04\,\AA, which increased the number of wavelength points from 25 in the case of non-equidistant spacing to 68 after resampling. 

In order to infer the LOS velocities from the individual spectrum in each pixel, we fit a second-order polynomial to the line core. Therefore, we determine the global minimum of the line and fit a second-order polynomial with a width around the line-core of $0.72$\,\AA\ for the H$\alpha$ line and $0.24$\,\AA\ for the NIR \mbox{Ca\,\textsc{ii}}. From the resulting parameters of the polynomial fit, we calculate the line-core position of each spectrum which is used to  calculate the Doppler velocity at each pixel. In addition, we obtain the standard deviation $\sigma$ for each coefficient of the polynomial from the fitting, and  calculate the uncertainty of the fit line-core position with the propagation of uncertainties according to Carl Friedrich Gau\ss{}. The uncertainties for the LOS velocity values, which we calculate from the fit line-core position, are obtained with the same procedure.

To evaluate the method, we scrutinize the systematic uncertainty introduced by the number of wavelength points used for the polynomial fitting. We changed number of wavelength points by $\pm1$, which resulted in a 10\% velocity difference for the H$\alpha$ profiles and a 12\% difference for the NIR  \mbox{Ca\,\textsc{ii}} profiles.

We inferred the magnetic field values from the HMI line-of-sight magnetograms. The uncertainties of these maps are about $\pm7$\,G produced by photon noise, as reported by \citet{Couvidat2016}. \citet{Kleint2017} studied the uncertainties of the polarization signal related to the photon noise from the IBIS instrument with about 1\% for the NIR \mbox{Ca\,\textsc{ii}} line. In contrast, we estimate the noise in the IBIS Stokes-$V$ maps with about 7--10\% of the intensity value by examining a quiet region in the maps.

\begin{figure*}[t]
\center
\includegraphics[width=1\textwidth]{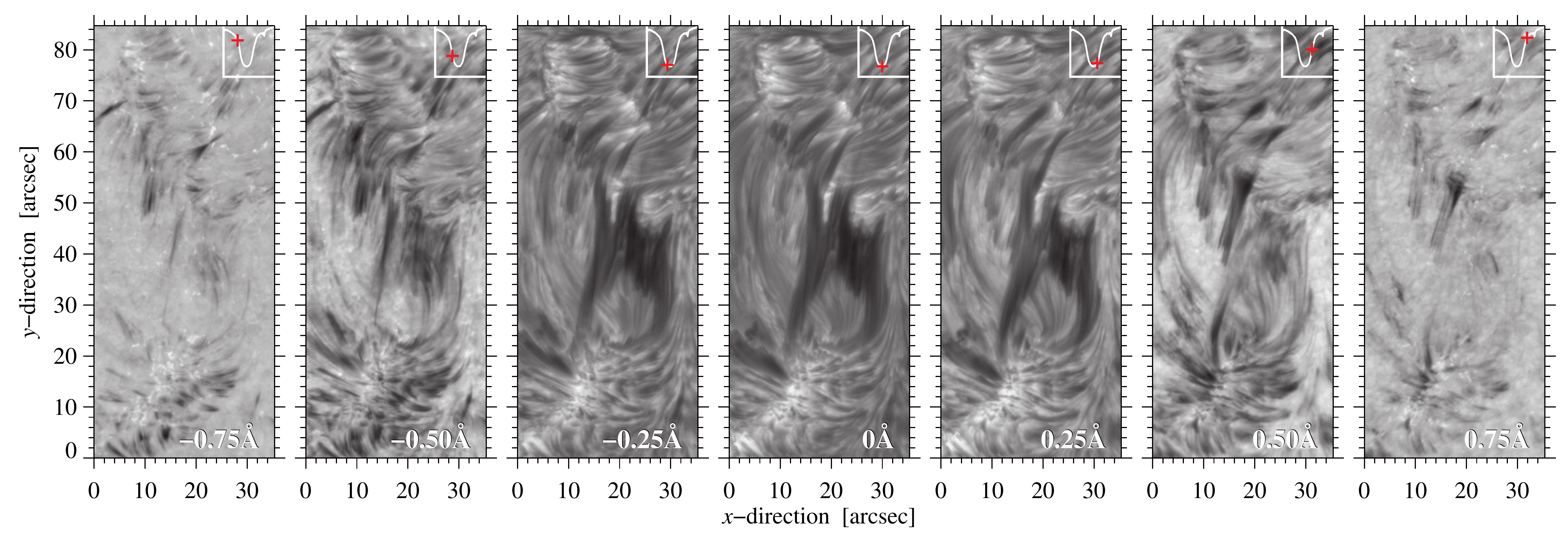}
\caption{Line scan of H$\alpha$ at 17:22:08~UT during best seeing conditions\protect\footnotemark. The scans include the line-wing positions H$\alpha \pm 0.75$\,\AA, H$\alpha \pm 0.5$\,\AA, and H$\alpha \pm 0.25$\,\AA\ and the line-core image ($\lambda_0 = 6562.86$\,\AA). The position within the line scan is indicated in the upper right corner of each panel.\label{FIG:ibis_best_6563}}
\end{figure*}

\begin{figure*}[t]
\center
\includegraphics[width=1\textwidth]{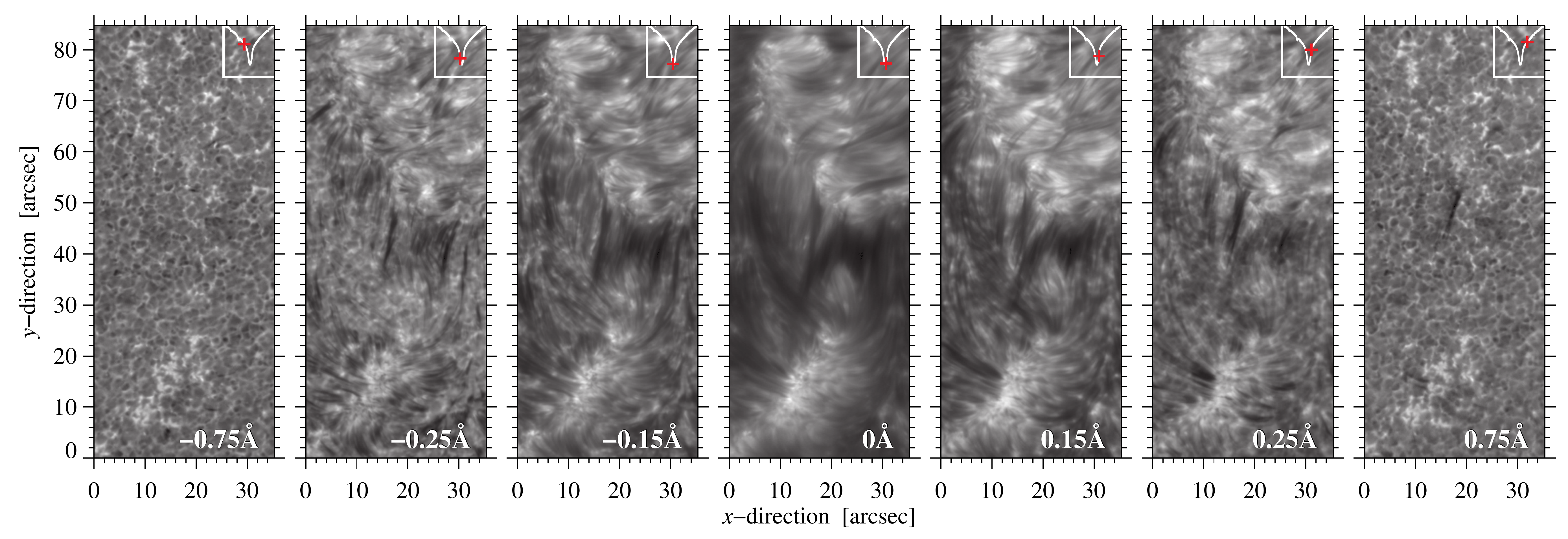}
\caption{Line scan of NIR~\mbox{Ca\,\textsc{ii}} at 17:21:42~UT during best seeing conditions\footref{note}. The scans show filtergrams at the wavelength positions \mbox{Ca\,\textsc{ii}} $\pm$ 0.75\,\AA, \mbox{Ca\,\textsc{ii}} $\pm$ 0.25\,\AA, and \mbox{Ca\,\textsc{ii}} $\pm$ 0.15\,\AA\ and the line-core filtergram ($\lambda_0 = 8542.21$\,\AA). The line core is located in the chromosphere, whereas the line wing filtergrams refer to successively lower regions in the
solar atmosphere.\label{FIG:ibis_best_8542}}
\end{figure*}

In addition, we interpolated the spectrum with a wavelength sampling of 0.01\,\AA. With this, we obtain images at H$\alpha \pm 0.75$\,\AA, H$\alpha \pm 0.5\,$\AA, and H$\alpha \pm 0.25$\,\AA\ (Fig.~\ref{FIG:ibis_best_6563}) and at NIR \mbox{Ca\,\textsc{ii}} $\pm$ 0.75\,\AA, \mbox{Ca\,\textsc{ii}} $\pm$ 0.25\,\AA, and \mbox{Ca\,\textsc{ii}} $\pm$ 0.15\,\AA\ (Fig.~\ref{FIG:ibis_best_8542}).

\section{Results} \label{sec:results}

\subsection{Classification of the filament system}

The main difference between AFSs and active region filaments is that a classical AFS connects the two opposite polarities of an EFR, whereas an active region filament resides above the PIL. In the present study, the magnetic configuration of the filament system is more complex so that a classification based on these two types is not trivial. Therefore, we discuss in this section the characteristics of the active region and the filaments therein.

Moving magnetic features are located at the periphery of the penumbra of the leading sunspot. They are often related to the moat flow and are a sign of sunspot decay. The MMFs are located close to the northern footpoint region of the filament system moving away from the sunspot. The other footpoint in the southern part is rooted in a decaying plage region, which already existed when active region NOAA 11658 rotated into view on 2013~January~13. The magnetic flux of the decaying plage spreads along the border of a supergranular cell.

\footnotetext{\label{note}A time-lapse movie referring to  Figs.~\ref{FIG:ibis_best_6563} and \ref{FIG:ibis_best_8542} is available in the online material.}

The AFS has a length of about 25--30\,Mm and a width of about 20\,Mm. The width of  individual filaments is about 3--4\,Mm. They have a lifetime of more than 60\,minutes and the whole system lives for several days, as seen in the AIA observations.  The filament system is very dynamic and changes its entire appearance during the two hours of observations. The magnetic configuration in this region is challenging. The positive footpoint is located close to the MMFs, where also negative polarity patches are in close range. Nonetheless, the filament system connects the positive polarity of the MMFs with the negative polarity at the southern footpoint along the border of a supergranule. 

Despite the longer lifetime of the individual arch filaments, which indicates that the system is more stable, the main characteristics of the filament system are similar to those of an classical AFS, that is, length, width, and number of the arch filaments \citep{Bruzek1967, GonzalezManrique2018}.  Nonetheless, the term AFS was defined as a system of small-scale filaments connecting newly emerged flux of opposite polarities that cross the PIL \citep{Bruzek1967}. In our case, the filament system is formed between MMFs of the decaying sunspot and slowly decaying flux along the border of a supergranular cell outside of the active region. The PIL itself is difficult to determine because of the complicated magnetic configuration but the filament system very likely crosses the PIL. Because of the divergence from the original definition, we call the system in the present study an ``extended AFS''.

In the following sections, we analyze the properties of the extended AFS and compare them to a classical AFS. In particular, we describe the morphological appearance, the temporal evolution, the LOS velocities, and the magnetic configuration.

\subsection{Morphology of the extended arch filament system} \label{sect:morph}

The FOV of ROSA shows the extended AFS and even small parts of the penumbra in the upper right corner (Fig.~\ref{FIG:ROSA}). In the G-band image, we recognize the common detail-rich granulation pattern of the photosphere with bright points in the intergranular lanes (Fig.~\ref{FIG:ROSA}a). The \mbox{Ca\,\textsc{ii}\,K} line (Fig.~\ref{FIG:ROSA}b) displays the inverse granular pattern, appearing as a dark central region surrounded by narrow bright lanes. In the \mbox{Ca\,\textsc{ii}\,K} line, regions with strong concentrations of magnetic field appear bright. The upper part of the image between coordinates (20\arcsec--\,59\arcsec, 35\arcsec--\,59\arcsec) is much brighter compared to the surrounding inverse granulation. The reason for this is likely the proximity to the sunspot, which resides in the upper right corner just outside of the FOV, and the  magnetic features around the sunspot. The bright region extends toward the lower part in a narrow strip at about the middle axis of the image, which is covering the position of the extended AFS in Fig.~\ref{FIG:ROSA}c. This panel represents the chromosphere in H$\beta$, where the extended AFS is visible. The images exhibit signatures of granulation, which result from the broad wings of the interference filter, which allows continuum radiation to pass. The whole image shows filamentary structures belonging to the surrounding fibrils. The extended AFS is located in the middle of the image (Fig.~\ref{FIG:ROSA}c) and appears as a dark structure. This extended AFS has not the classical elongated appearance but possesses a more clumpy and cloudy structure, so that we cannot distinguish single threads. An elongated filamentary structure is located at the left side of the extended AFS. The extended AFS connects MMFs just outside the penumbra and the southern footpoint of the system. G-band bright points are located in the northern and southern footpoints (Fig.~\ref{FIG:ROSA}a and~c).

In H$\alpha$, the extended AFS appears very similar to the structures in  H$\beta$ but we can distinguish single threads (Fig.~\ref{FIG:ibis_best_6563}, line-core). A detailed comparison of the H$\alpha$ and H$\beta$ line is provided in Sect.~\ref{sec:comparison}. The H$\alpha$ line-core filtergram shows the extended AFS in the middle of the FOV. The filamentary structure of the chromospheric fibrils is typical for the H$\alpha$ line. The footpoints at (20\arcsec, 65\arcsec) and (10\arcsec, 18\arcsec) appear bright. At H$\alpha \pm 0.75$\,\AA, we recognize dark mottles at both footpoints but with different shapes for both sides (Fig.~\ref{FIG:ibis_best_6563}). \citet{Bray1974} reported that mottles have their maximal visibility at H$\alpha \pm 0.75$\,\AA. Furthermore, bright structures are located at the footpoints. In the filtergrams at H$\alpha \pm 0.50$\,\AA, more and more filamentary structures become visible (Fig.~\ref{FIG:ibis_best_6563}). First, the elongated filamentary structure on the left side of the AFS becomes visible. However, the clumpy main structure  becomes visible in the filtergrams closer to the line core. In the regions H$\alpha\pm 0.50$\,\AA, the structures appear different, which may indicate up- or downstreaming flows in the loop tops and footpoints, respectively. The filtergrams at H$\alpha \pm 0.25$\,\AA\ are very similar to the line-core filtergrams.

The extended AFS in the NIR~\mbox{Ca\,\textsc{ii}} line-core filtergrams appears slightly different (Fig.~\ref{FIG:ibis_best_8542}) compared to H$\alpha$: (1) The NIR~\mbox{Ca\,\textsc{ii}} line shows different structures as the ROSA \mbox{Ca\,\textsc{ii}\,K} line at $\lambda3934$\,\AA\ (Fig.~\ref{FIG:ROSA}b). This can be explained by assuming that the NIR~\mbox{Ca\,\textsc{ii}} line core is located higher in the chromosphere than the \mbox{Ca\,\textsc{ii}\,K} line. Moreover, the filter of the \mbox{Ca\,\textsc{ii}\,K} line is a broad-band filter, which does not show the upper chromospheric features. (2) The filamentary appearance of the chromosphere is similar to H$\alpha$. (3) The two footpoints appear very bright. (4) The footpoints coincide with those in H$\alpha$. (5) The size of the filamentary structures is much smaller compared to H$\alpha$.

At the locations NIR~\mbox{Ca\,\textsc{ii}} $\pm$ 0.15\,\AA, the arch filament system is still visible but patterns similar to inverse granulation shimmer through the background. Only $\pm0.25$\,\AA\ away from the line core, the morphology of the lower layers changes significantly. Here, most structures resemble the inverse granulation. The bright features at the footpoint location are related to strong magnetic field concentrations. Furthermore, thin dark structures, similar to mottles originate in the bright regions, which may be related to the arch filament in higher atmospheric layers. At \mbox{Ca\,\textsc{ii}} $\pm$ 0.75\,\AA, the images show inverse granulation and bright regions associated with magnetic fields similar to the \mbox{Ca\,\textsc{ii}\,K }line, as in the ROSA data (Fig.~\ref{FIG:ROSA}b). The \mbox{Ca\,\textsc{ii}\,K} line originates from the upper photosphere and lower chromosphere and forms deeper than the NIR~\mbox{Ca\,\textsc{ii}} line. In any case, the spectral lines cover a large range of heights in the atmosphere \citep{Stix2004} and structures overlap.

\begin{figure}
\center
\includegraphics[width=1\columnwidth]{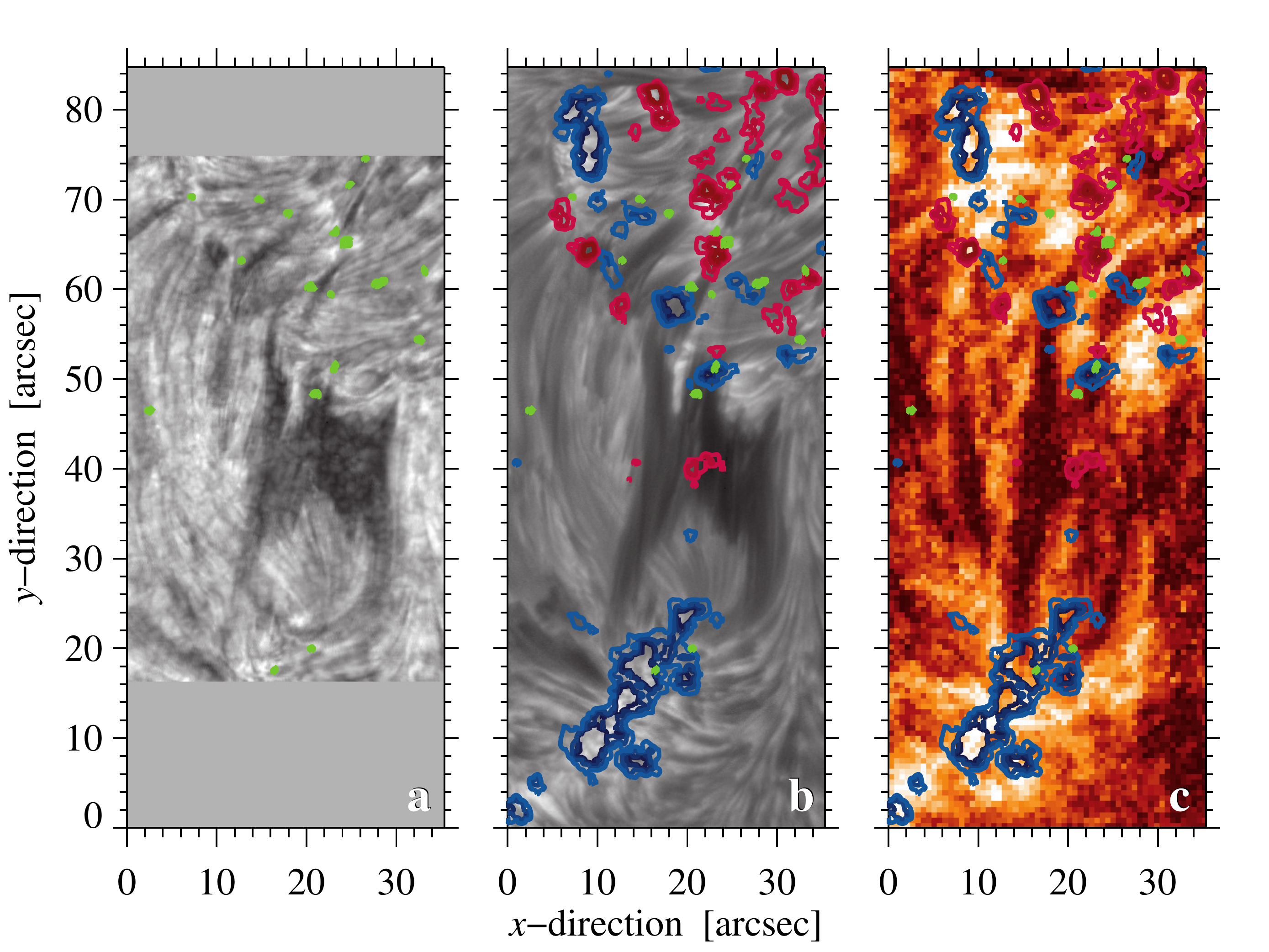}
\caption{Comparison of the extended AFS observed with different instruments. (a)~ROSA H$\beta$ image with contours of G-band bright points (green). (b)~IBIS H$\alpha$ line-core filtergram with contours of the positive (red) and negative (blue) LOS magnetic field from HMI between $\pm50$\,G and  $\pm200$\,G in steps of 50\,G, where darker colors indicate stronger magnetic fields. (c)~NAFE-enhanced AIA image.\label{FIG:footpoints_overview}}
\end{figure}

\begin{figure*}[t]
\center
\includegraphics[width=1\textwidth]{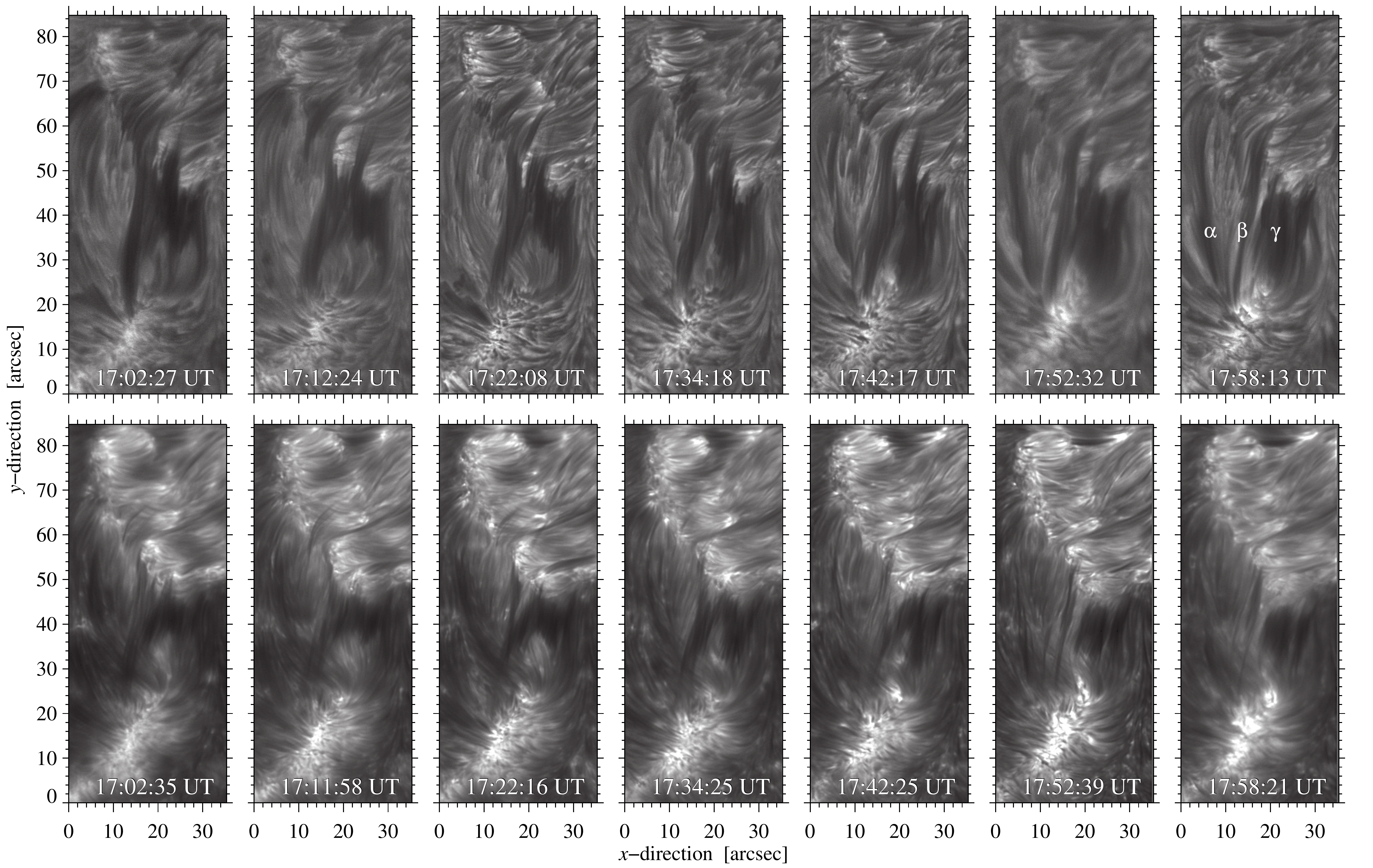}
\caption{Temporal evolution of the extended AFS in the chromosphere in selected H$\alpha$ line-core filtergrams (top) and in NIR~\mbox{Ca\,\textsc{ii}} line-core filtergrams (bottom) for a time-series starting at 17:02\,UT until 17:58\,UT on 2013~January~20. The labels $\alpha$, $\beta$, and $\gamma$ indicate three individual arch filaments, which are discussed in the text.\label{FIG:ibis_temp}}
\end{figure*}

\subsection{Comparison of H$\alpha$ and H$\beta$} \label{sec:comparison}

The multi-instrument observations provide the opportunity to compare IBIS 
H$\alpha$ line-core images with H$\beta$ images of ROSA 
(Fig.~\ref{FIG:footpoints_overview}). In addition, we compare both wavelength 
regions with AIA images. The IBIS FOV is used as a reference, and regions beyond 
the smaller ROSA FOV are displayed in uniform gray 
(Fig.~\ref{FIG:footpoints_overview}). A cursory glance reveals very similar 
structures, meaning that, both chromospheric absorption lines show the extended 
AFS in its largest dimension. However, a closer inspection shows notable 
differences in the fine structure. The filamentary background in H$\beta$ is not 
as dominant as in H$\alpha$, and structures in the background are brighter. In 
the H$\alpha$ images, both footpoints appear much brighter. Furthermore, the 
fine structure is distinguishable as single threads, which is not the case for 
the extended AFS in H$\beta$, where it appears more like a cloudy structure. The 
endpoints of single filaments are more distinct in H$\alpha$ than in H$\beta$, 
and also some fine structures are more clearly apparent in H$\alpha$. 
Nonetheless, H$\beta$ is a good alternative in the blue wavelength range to 
observe filaments, AFS, or other chromospheric absorption structures.

In comparison, the extended AFS in the AIA image appears different, but some filamentary structures are still recognizable (Fig.~\ref{FIG:footpoints_overview}c), while the clumpy structure in the middle is absent in the AIA images, leaving the arch filaments on both sides clearly separated. Furthermore, we present in Fig.~\ref{FIG:footpoints_overview} the location of the  photospheric G-band bright points (green contours) from the ROSA G-band images (Fig.~\ref{FIG:ROSA}a). Many G-band bright points are encountered in the vicinity of the sunspot, especially in proximity to the northern footpoint region. They are located close to both footpoints of the single filaments of the extended AFS, which corresponds to regions with relatively strong magnetic field concentrations.

\subsection{Temporal evolution of the extended arch filament system}\label{sect:tempev}

We have continuous H$\beta$ observations (ROSA imaging) of the extended AFS for about an hour between 16:52:03\,--\,17:57:46~UT as well as H$\alpha$ and NIR~\mbox{Ca\,\textsc{ii}} (IBIS spectroscopy and spectropolarimetry). Because of the similarity in H$\alpha$ and H$\beta$, we discuss the temporal evolution of the extended AFS in both wavelengths together and compare this to the temporal evolution in NIR~\mbox{Ca\,\textsc{ii}}.

In the upper row of Fig.~\ref{FIG:ibis_temp}, we show seven selected points in time (approximately 10\,minutes apart)  of the H$\alpha$ line-core image tracing the filament's evolution. The extended AFS appears as a clumpy structure at the beginning of the time-series (Fig.~\ref{FIG:ibis_temp}, upper row at 17:02\,UT). The clumpy structure covers most of the FOV. The rest of the image contains fibrils and some filamentary structures at the upper-left border of the image. In time-lapse movies (online movie referring to Figs.~\ref{FIG:ibis_best_6563} and \ref{FIG:ibis_best_8542}), plasma from the northern footpoint region moves toward the southern footpoint, passing the  main body of the filament. In the movies, the filamentary structures of the main body become more and more separated (Fig.~\ref{FIG:ibis_temp}, arch filaments $\beta$ and $\gamma$) and another loop builds up to the left of the extended AFS (Fig.~\ref{FIG:ibis_temp}, arch filament $\alpha$) at around 17:42\,UT. The clumpy structure first extends in width until it reaches a maximum at around 17:15:56~UT. Next, the separation is visible in Fig.~\ref{FIG:ibis_temp} until the arch filaments are completely separated at 17:58:16\,UT. At the end of the time-series, we identify two clearly separated elongated arch filaments in the extended AFS (Fig.~\ref{FIG:ibis_temp}, arch filaments $\alpha$ and $\beta$) and some more filamentary structures connecting the two footpoint regions (Fig.~\ref{FIG:ibis_temp}, arch filament $\gamma$).

\begin{figure*}
\center{
\includegraphics[width=1\textwidth]{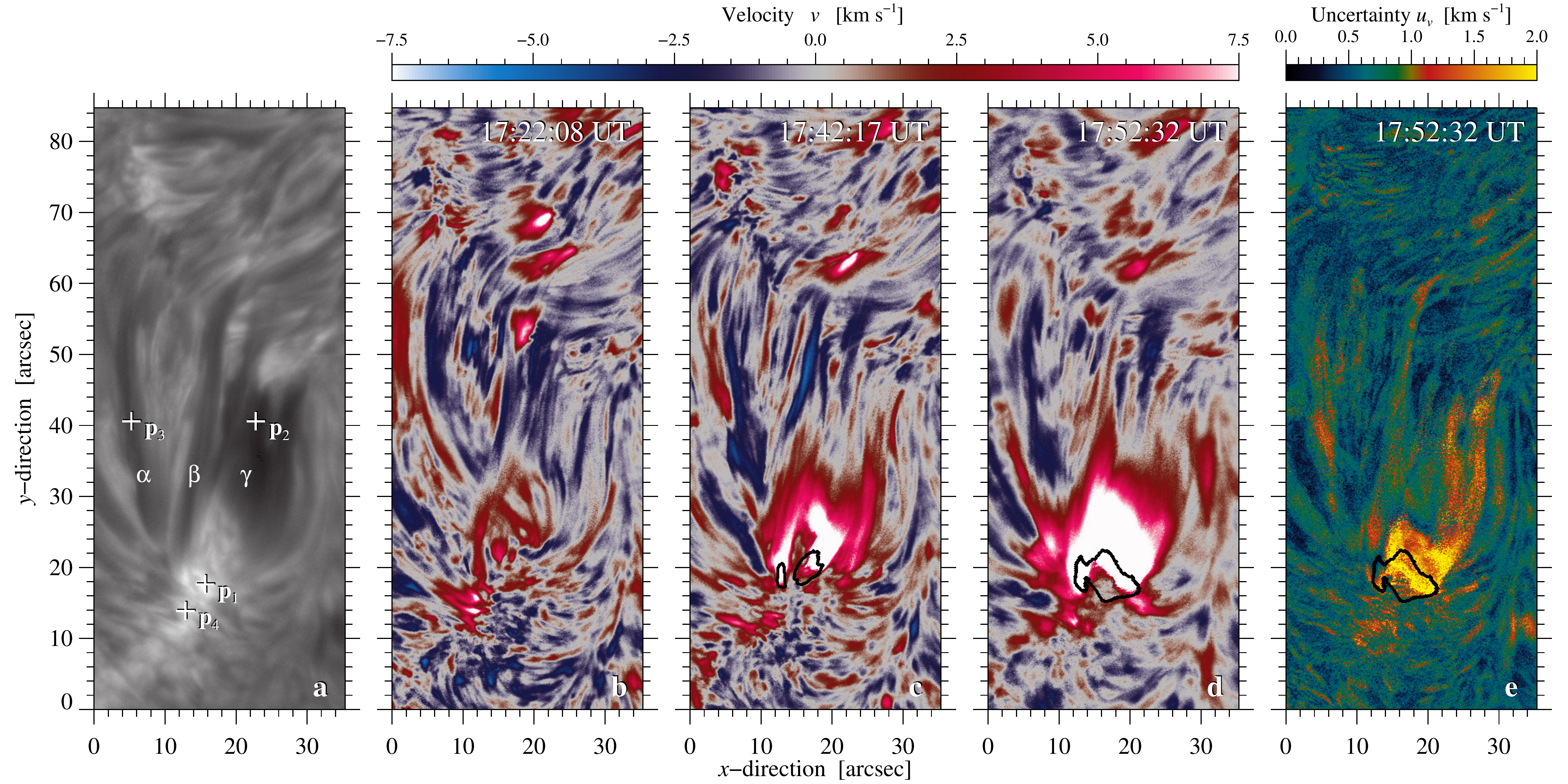}
\caption{(a) H$\alpha$ line-core image and Doppler velocity maps (b) at 17:22:08\,UT, (c) at 17:42:17\,UT, and (d) at 17:52:32\,UT. All velocity maps are clipped between $\pm7.5$\,km\,s$^{-1}$, i.e., the saturated white regions denote velocities exceeding this limit. The points p$_1$\,--\,p$_4$ refer to regions, which we examine in detail. The black contours show the regions with two spectral components indicating very strong redshifts. The labels $\alpha$, $\beta$, and $\gamma$ indicate three individual arch filaments, which are discussed in the text. (e) Map of the uncertainties $u_v$ of the LOS velocities from the line-core fits exemplary for the LOS map at 17:52:32\,UT \label{FIG:ibis_phys_6563}}}
\end{figure*}

The clumpy appearance of the extended AFS is possibly an optical effect, because the single threads are very close to each other. Especially, this effect is apparent in H$\beta$. Furthermore, bright structures at the northern footpoint region shimmer through the filamentary structure of the clumpy arch filament (Fig.~\ref{FIG:ibis_temp} at 17:34~UT). However, all arch filament structures are connected to the southern footpoint, which can be seen as the strong anchor of the AFS, where most plasma is transported. The time-lapse movies show that the evolution of the extended AFS is very fast and even during the short period of one hour the appearance of the arch filament changed significantly. The line-core images of the NIR~\mbox{Ca\,\textsc{ii}} line are displayed in the lower row of Fig.~\ref{FIG:ibis_temp}. The overall behavior of the extended AFS is the same in the H$\alpha$ and NIR~\mbox{Ca\,\textsc{ii}} lines.

At the end of the time-series, dominant downstreaming flows are present in the southern footpoint in the H$\alpha + 0.75$\,\AA\ line-wing filtergram. This leaves the impression that material is moving toward this footpoints. Such downstreaming flows at the footpoints are typical for classical AFS and are also present in the extended AFS. We analyze these strong downstreaming flows in Sect.~\ref{sect:ibis_LOS}.

In the time-lapse movie of the AIA images (online movie referring to Fig.~\ref{FIG:ibis_mag_align}), a brightening appears in the newly developed arch filament at around 17:20\,UT at the left side of the extended AFS (Fig.~\ref{FIG:ibis_temp}, arch filament $\alpha$), which is moving from the northern footpoint region to the southern footpoint along the arch filament. After this brightening in the AIA images, we recognize a variation in the spectra as a redshift of the spectra in H$\alpha$ as well as in NIR~\mbox{Ca\,\textsc{ii}} related to the southern footpoint. Thus, the plasma is flowing most likely downward along the loops. These spectra will be analyzed in the following section.

\begin{figure*}
\includegraphics[width=1\textwidth]{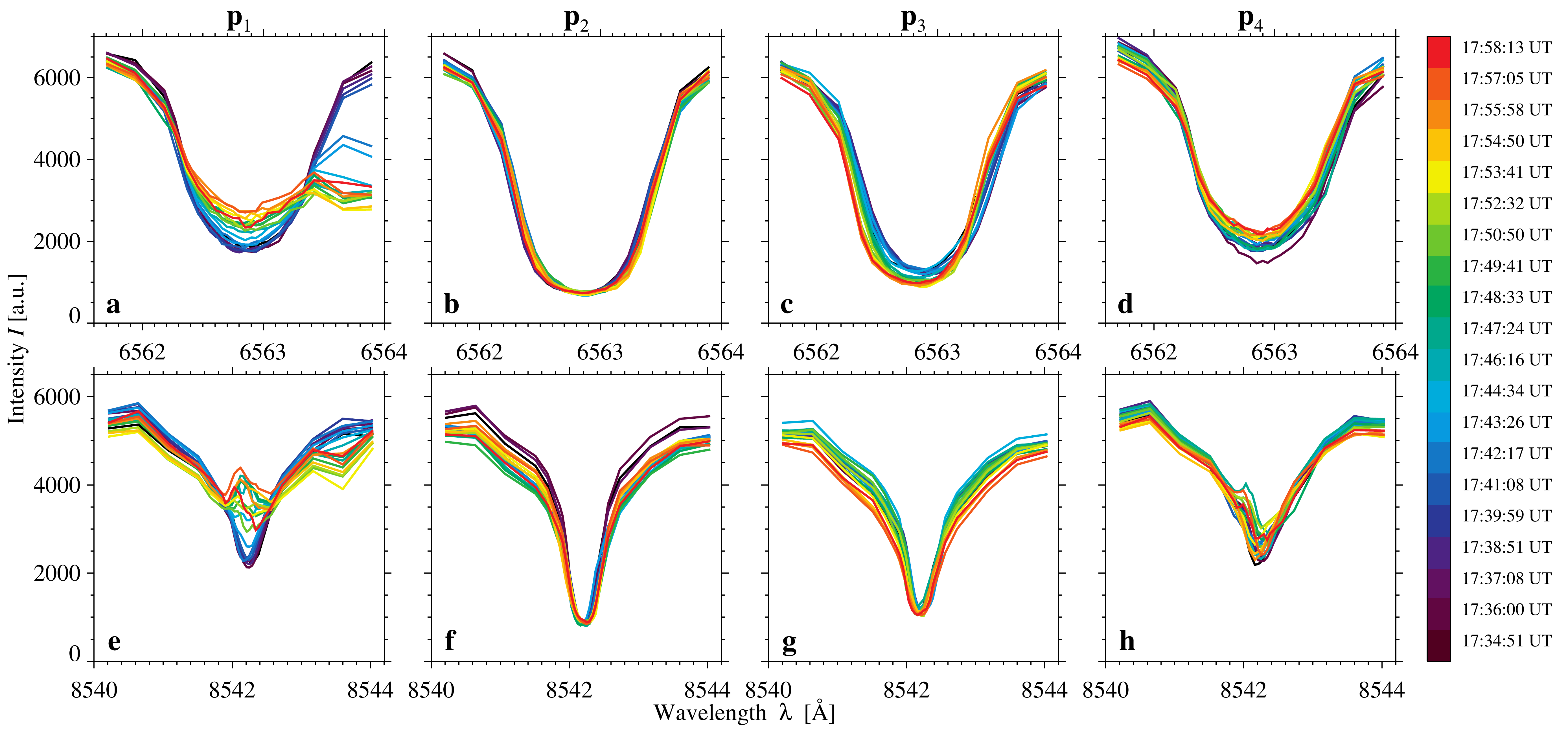}
\caption{Temporal evolution of the H$\alpha$ (top) and   
    NIR~\mbox{Ca\,\textsc{ii}} (bottom) observed spectra at points
    p$_1$\,--\,p$_4$ (left to right). The color of the spectral profiles
    corresponds to the observing time given by the scale on the right.\label{FIG:ibis_spectra}}
\end{figure*}

\subsection{Line-of-sight velocities of the extended arch filament system}\label{sect:ibis_LOS} 

We use the resampled IBIS data cubes (Sect.~\ref{sect:prefilter}) to determine the LOS velocity. First, we determine the average quiet-Sun profile, where we excluded very dark and bright regions, as well as regions with strong red- and blueshifts. Second, we fit for each image a second-order polynomial to the line core of the averaged quiet-Sun profile to determine the wavelength at rest.  Afterward, the wavelength shift is inferred by line-core fitting of the individual IBIS profiles, as described in Sect.~\ref{sect:prefilter}. Applying the Doppler formula, using $\lambda_0 = 6562.81$\,\AA\ as the laboratory H$\alpha$ wavelength \citep{Moore1966},  yields the LOS velocities shown in the maps of Fig.~\ref{FIG:ibis_phys_6563} for different observing times. Blue colors indicate negative line shifts and red colors indicate positive line shifts.

\begin{figure*}
\center{
\includegraphics[width=1\textwidth]{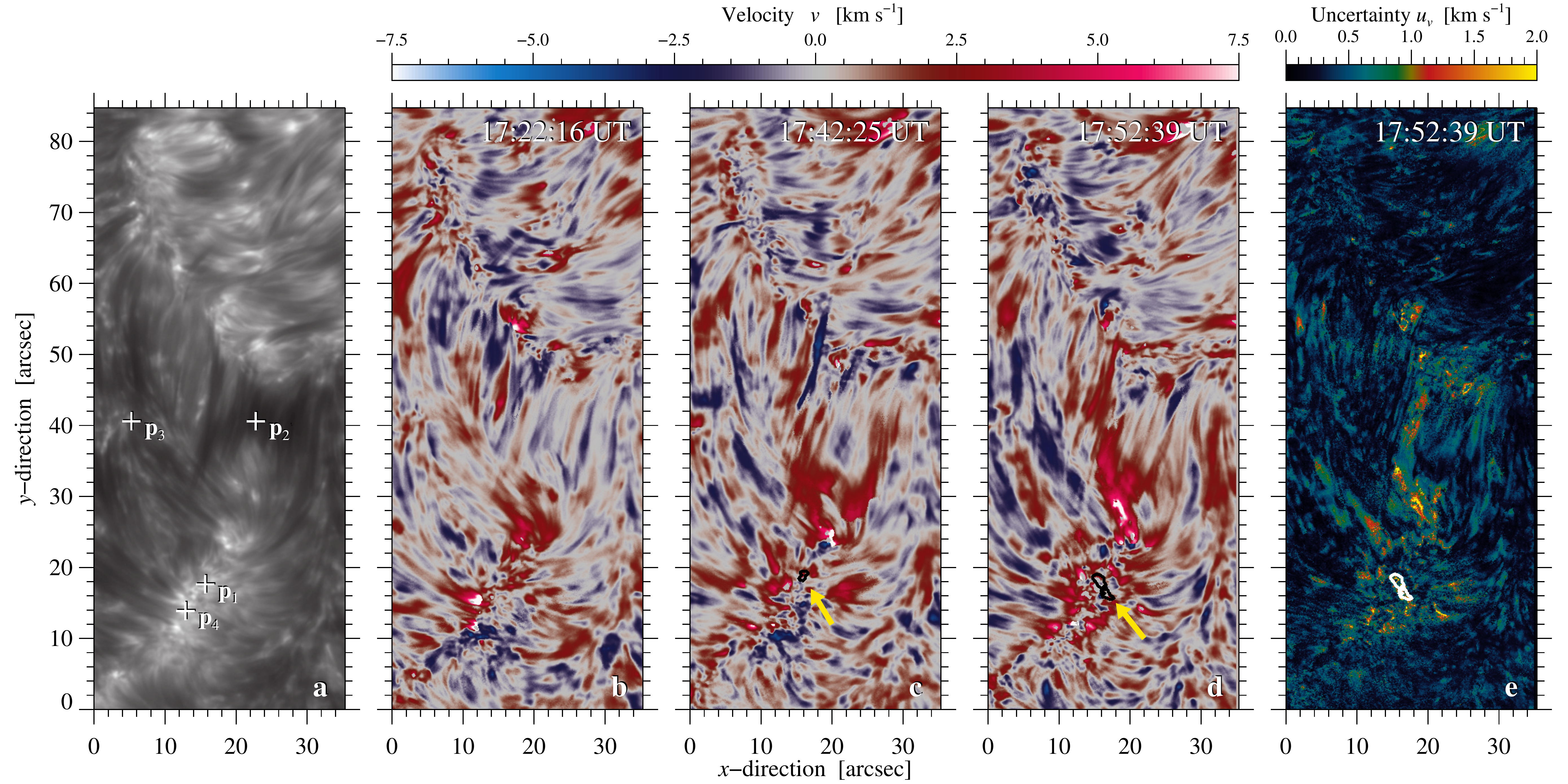}
\caption{(a) NIR~\mbox{Ca\,\textsc{ii}} line-core image and 
    Doppler velocity maps (b) at 17:22:16\,UT, (c) at 17:42:25\,UT, and (d) at 17:52:05\,UT. All velocity maps are clipped between $\pm7.5$\,km\,s$^{-1}$. The points p$_1$\,--\,p$_4$ denote regions, which we examine in detail. The black contours, indicated by a yellow arrow, show the regions with two spectral components indicating very strong redshifts. (e) Map of the uncertainties $u_v$ of the LOS velocities from the line-core fits exemplary for the LOS map at 17:52:32\,UT.\label{FIG:ibis_phys_8542}}}
\end{figure*}

\citet{Bruzek1967, Bruzek1969} describes that individual arch filaments rise in the atmosphere, which is indicated by blueshifts at the loop tops. At the footpoints the plasma drains along the arches as indicated by redshifts. Because of this, we associate the blueshifts with upflows and the redshifts with downflows, even though we do not observe at disk center, and we are aware of the influence of differential rotation (with $\mu \approx 0.9$) and LOS projection, which may lead to variations across the FOV.

\begin{SCfigure*}
\includegraphics[width=0.66\textwidth]{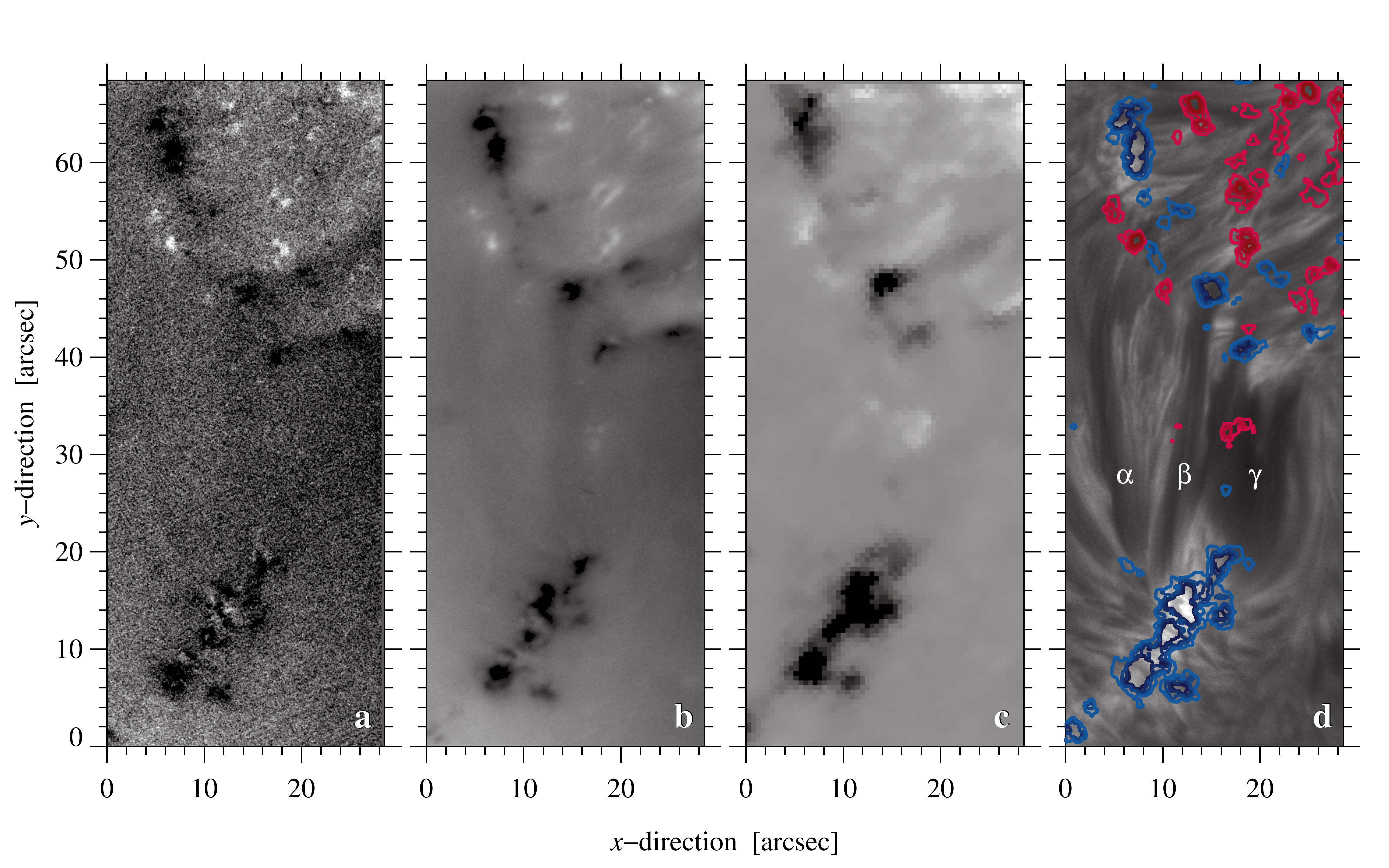}
\caption{Magnetic field of the extended AFS. (a) Single-integration IBIS Stokes-$V/I_\mathrm{c}$ map derived from the  NIR~\mbox{Ca\,\textsc{ii}} line at 17:52:39\,UT and (b) one-hour average of IBIS Stokes-$V/I_\mathrm{c}$ maps scaled between $V/I_\mathrm{c} = \pm 0.005$. (c) Two-hour average HMI magnetogram clipped between $\pm150$\,G, showing the positive (white) and negative (black) magnetic polarities.  (d) H$\alpha$ image at 17:52:39\,UT with contours of the positive (red) and negative (blue) magnetic field between $\pm50$\,G and  $\pm200$\,G in steps of 50\,G, where darker colors indicate stronger magnetic fields. The contours were derived from a single HMI magnetogram at 17:36:00\,UT. The labels $\alpha$, $\beta$, and $\gamma$ indicate three individual arch filaments, which are discussed in the text.\protect\footnotemark \label{FIG:ibis_mag_align}}
\end{SCfigure*}

At 17:22:08\,UT (Fig.~\ref{FIG:ibis_phys_6563}b), we already see strong downstreaming flows at the footpoints, and in the central part (see Fig.~\ref{FIG:ibis_phys_6563}a, arch filament $\gamma$) of the extended AFS we detect upstreaming flows, which is expected for the cool loops of a classical AFS. The newly formed arch filament on the left of the extended AFS exhibits predominant downflows (see Fig.~\ref{FIG:ibis_temp}, arch filament  $\alpha$). Based on the AIA time-lapse movies (online movie referring to Fig.~\ref{FIG:ibis_mag_align}), we can relate this structure to the bright structure in the AIA images, which appeared at the same time and location. Slightly blueshifted flows are notable in the map, which reach maximum velocities of up to ($5.3\pm0.9$)\,km\,s$^{-1}$. The downflows reach maximum velocities of up to ($9.8\pm0.6$)\,km\,s$^{-1}$ around the footpoints. In this region, the LOS velocities (redshifts) in the southern footpoint increase with time. At 17:42:17\,UT (Fig.~\ref{FIG:ibis_phys_6563}c), the downstreaming flows in this region reach values of up to ($18.7\pm0.7$)\,km\,s$^{-1}$. The strong downflows in the aforementioned new arch filament $\alpha$ are moving toward the southern footpoint. The peak velocity of the upflows did not change significantly. At 17:52:32\,UT (Fig.~\ref{FIG:ibis_phys_6563}d), the velocity of the downstreaming flows increased again, reaching values of up to $(22.8\pm0.6)$\,km\,s$^{-1}$. The lower part of the FOV shows predominantly redshifted structures. In addition, the filamentary structure $\alpha$ on the left side of the FOV shows stronger downflows located closer to the southern footpoint. The maximum upflows during this time decrease to $(3.9\pm1.1)$\,km\,s$^{-1}$. However, by this time, the upflows in the central part of the extended AFS (see Fig.~\ref{FIG:ibis_phys_6563}a, arch filament $\gamma$) are completely replaced by downflows. 

In addition to the velocity maps, we provide a map displaying the uncertainties (Fig.~\ref{FIG:ibis_phys_6563}e) for the map at 17:52:32\,UT (Fig.~\ref{FIG:ibis_phys_6563}d). The value of the uncertainty corresponding to the 99th percentile is 1.79\,\,km\,s$^{-1}$. The number of poor fits is 0.01\%, meaning that, the line core cannot be properly fit with a parabola over the given range.

For comparison, we present in Fig.~\ref{FIG:ibis_spectra} the temporal evolution of the spectra at four selected points, which are marked in Fig.~\ref{FIG:ibis_phys_6563}a. The first point p$_1$ is in the middle of the southern footpoint. Here, we see the most significant changes in the spectrum (Fig.~\ref{FIG:ibis_spectra}a). The spectra become more and more redshifted with time, and a second component appears in the H$\alpha$ spectrum. Because of the sparse wavelength sampling in this part of the spectrum, about 48\,km\,s$^{-1}$ represents a rough estimate for the second component's velocity, which is twice as much as for the first component. The second component starts to appear at around 17:42\,UT and disappears again at around 17:53\,UT. The location of the profiles, which show line gaps and a second component are outlined in Fig.~\ref{FIG:ibis_phys_6563} with the black contours, which increases over time. Point p$_4$ (Fig.~\ref{FIG:ibis_spectra}d) is located in proximity to p$_1$ in the southern footpoint. Here, the spectrum remains unchanged. Point p$_2$ (Fig.~\ref{FIG:ibis_spectra}b) marks a location with strong absorption in the central part $\gamma$ of the extended AFS. Here, a small blueshift appears with time, which was also visible in the LOS velocity maps of Fig.~\ref{FIG:ibis_phys_6563}. The blueshift decreases over time and becomes a redshift at the end of observations. Point p$_3$ (Fig.~\ref{FIG:ibis_spectra}c) labels the newly formed arch filament $\alpha$ on the left side, that is, the location, where a dark feature appears during the temporal evolution of the extended AFS. During the temporal evolution of the spectra, we determine an increasing redshift for this region.

The Doppler velocities of the NIR~\mbox{Ca\,\textsc{ii}} profiles are derived in the same way as the H$\alpha$ profiles, that is, with line-core fitting (Sect.~\ref{sect:prefilter}). The reference wavelength was $\lambda_{0, \mathrm{ref}} = 8542.14$\,\AA\ \citep{Moore1966} for the quiet-Sun profile. At locations with line-core emission (see Fig.~\ref{FIG:ibis_spectra}e), we set the velocity to zero. The velocities at three different times are shown in Fig.~\ref{FIG:ibis_phys_8542}. At 17:22:16\,UT (Fig.~\ref{FIG:ibis_phys_8542}b), the footpoints are clearly visible showing high redshifts. In addition, a region in the upper right corner from the direction of the sunspot is dominated by downflows. These downflows reach at that time LOS velocities of up to ($12.7\pm0.2$)\,km\,s$^{-1}$. Small upflows are present in most regions and reach LOS velocities of only up to ($4.4\pm0.3$)\,km\,s$^{-1}$. At 17:42:25\,UT, the downflows close to the southern footpoint increase reaching values of up to ($17.8\pm0.5$)\,km\,s$^{-1}$. However, also downflows increase in the upper right corner, which become visible as the saturated structure. Furthermore, the upflows in the entire FOV increase and reach values of up to ($6.7\pm1.1$)\,km\,s$^{-1}$. Strong upflows develops at the location of the loop top of one arch filament, indicating the rise of the arch filament. At 17:52:39\,UT, the strong upflow structure in the central filamentary structure disappears and is replaced by downflows at (20\arcsec, 40--60\arcsec) in Fig.~\ref{FIG:ibis_phys_8542}d. The redshifted regions indicate an increase of the downflows at the southern footpoint. 

We display the map of uncertainties (Fig.~\ref{FIG:ibis_phys_8542}e) for the map at 17:52:32\,UT (Fig.~\ref{FIG:ibis_phys_8542}d) to evaluate the uncertainties of the LOS velocity. The value of the uncertainty corresponding to the 99th percentile is 1.13\,\,km\,s$^{-1}$. The number of poor fits is 0.1\%, in particular, where the line core cannot be properly fit with a parabola over the given range.

The computed velocities in this region have to be carefully interpreted because the spectra show a prominent shoulder in the red line-wing, hampering the exact determination of LOS velocities by line-core fitting (see Fig.~\ref{FIG:ibis_spectra}e).  Here, we present spectra of point p$_1$ in the NIR~\mbox{Ca\,\textsc{ii}} line as a function of time. We recognize inner line-core emission. Furthermore, we see as before in the H$\alpha$ line that a second component develops with LOS velocities, which can be roughly estimated with a velocity of about 52\,km\,s$^{-1}$.  The second component in the NIR~\mbox{Ca\,\textsc{ii}} line develops about five minutes later than in H$\alpha$. The region, where the second component appears in the spectral profiles, is marked by black contours in Fig.~\ref{FIG:ibis_phys_8542}. The region is much smaller than for H$\alpha$ but also grows over time. The spectral profiles of point p$_4$, close to point p$_1$, shows also line-core emission but not as strong as for point p$_1$. For points p$_2$ and p$_3$, we do not see any unusual spectral features but we recognize a red- and blueshift, respectively, at these points (see Fig.~\ref{FIG:ibis_phys_8542}).

\subsection{Magnetic field of the extended arch filament system}\label{sect:IBISmag}

\footnotetext{\label{note2}A time-lapse movie referring to  Fig.~\ref{FIG:ibis_mag_align} is available in the online material including the temporal evolution in the AIA wavelength band.}

We analyzed IBIS Stokes-$V$ profiles of the NIR~\mbox{Ca\,\textsc{ii}} line, which contain information about the circular polarization. The linear polarization signals were too weak to analyze them more analytically. Furthermore, single Stokes-$V$ profiles are very noisy (Fig.~\ref{FIG:ibis_mag_align}a). Averaging single maps across the two-hour time-series decreases the noise significantly. This is justified because no major changes occurred in the magnetic field during this time period. The resulting Stokes-$V/I_\mathrm{c}$ map is shown in Fig.~\ref{FIG:ibis_mag_align}b. The corresponding two-hour average of HMI magnetograms is aligned with respect to the FOV of IBIS (Fig.~\ref{FIG:ibis_mag_align}c). For comparison, we present the H$\alpha$ line-core image in Fig.~\ref{FIG:ibis_mag_align}d. In addition, the contours of the magnetic field, based on one HMI map at 17:36:00\,UT, are superposed on the image. 

At this state of the analysis, we have no information about the strength of the chromospheric magnetic field because single-integration Stokes-$V$ maps are just too noisy for inversions. Hence, we resorted to simply comparing the appearance of the averaged IBIS Stokes-$V$ map to the average HMI magnetogram. The average HMI magnetogram shows the magnetic field in the photosphere, whereas the NIR~\mbox{Ca\,\textsc{ii}}~line reveals information about the LOS magnetic field in the chromosphere. Overall, both maps are very similar. The northern part of the FOV has mixed polarities, whereby the stronger fields have negative polarity. Here, the positive polarity is related to MMFs. In the single-integration IBIS Stokes-$V$ map as well as in the averaged map (Fig.~\ref{FIG:ibis_mag_align}a~and~b), we recognize more positive polarity flux in the upper right corner in proximity to the sunspot. Furthermore, both maps display a ``haze'' of negative polarity at the location of the arch filament $\beta$ (Fig.~\ref{FIG:ibis_mag_align}d), which is probably related to the weak magnetic field of the extended AFS in the chromosphere. This haze is also visible for positive polarities (Fig.~\ref{FIG:ibis_mag_align}b and c, upper right corner), and is maybe connected to the magnetic field of the sunspot.  Across two different atmospheric layers, the two magnetic field maps appear very similar. The southern part of the FOV is dominated by negative polarity. In the time-lapse movies, traces of positive polarity are visible in the predominant negative polarity, which signifies flux cancellation in this region.

Furthermore, we compare the H$\alpha$ line-core filtergram (Fig.~\ref{FIG:ibis_mag_align}d) with the contours of the HMI magnetic field. The elongated arch filaments of the extended AFS connect positive flux of the northern footpoint region, related to MMFs, with negative flux of the southern footpoint. We have positive flux in the middle of the extended AFS at the coordinates (18\arcsec, 34\arcsec) and a negative flux at coordinates (18\arcsec, 42\arcsec). We surmise that the central structure of the extended AFS emerged between these two polarities. This central structure is not visible in the AIA images. Thus, we conjecture that this structure resides lower in the atmosphere than the remainder of the extended AFS.


\section{Discussions} \label{sect:disc}

\begin{table*}[t]
\begin{center}
\caption{Comparison of the different properties of a ``classical'' AFS and the ``extended'' AFS.}
\begin{tabular}{lll}
\hline\hline
Properties &  classical AFS & extended AFS\rule[0mm]{0mm}{3mm}\\
\hline
Length & $\sim$30\,Mm{$^\dagger$} & 25--30\,Mm\rule[0mm]{0mm}{4mm}\\
Width (system) & $\sim$20\,Mm{$^\dagger$}  & $\sim$20\,Mm\rule[0mm]{0mm}{4mm}\\
Width (arch filament) & 1--3\,Mm{$^\dagger$} & $\sim$3\,Mm\rule[0mm]{0mm}{4mm}\\
Location & EFR & active region and surrounding\rule[0mm]{0mm}{4mm}\\
Connectivity & pores, small-scale flux & MMFs $\&$ network magnetic elements\rule[0mm]{0mm}{4mm}\\
Footpoints & footpoints drifting apart & MMFs move toward other footpoint \rule[0mm]{0mm}{4mm}\\
Loops & rise from below the photosphere & ascent starts in photosphere \rule[0mm]{0mm}{4mm}\\
Flux & emerging flux & decaying flux \rule[-2mm]{0mm}{6mm}\\
\hline
\end{tabular}
\label{tab:comparison}
\end{center}
\parbox{60mm}{
\vspace{-3mm}\hspace{27mm}\footnotesize{$^\dagger$}\citet{Bruzek1967, Bruzek1969}}
\end{table*}

This arch filament system is not a classical one, which typically connects positive and negative emerging flux above a young active region. In the present case, we have an AFS that connects positive polarity of MMFs in the vicinity of the sunspot penumbra with the remaining negative flux of the quiet Sun, which spreads  at the border of a supergranule. In Table~\ref{tab:comparison}, we compare the properties of a classical AFS with those of an extended AFS. In the temporal evolution of the AFS, we noted that the clumpy AFS evolves into a separated and elongated arch filament (Fig.~\ref{FIG:ibis_temp}, arch filament $\beta$) and some more filamentary structures connecting both footpoints (Fig.~\ref{FIG:ibis_temp}, arch filament $\gamma$). Furthermore, we observed the formation of another elongated arch filament during the observations (Fig.~\ref{FIG:ibis_temp}, arch filament $\alpha$). All these phenomena show that the chromosphere in the vicinity of the active region was very dynamic. In addition, we point out that the extended AFS does not belong to the superpenumbra of the sunspot, which is easily apparent in the H$\beta$ and H$\alpha$ images. In consequence, the inverse Evershed flow \citep{Evershed1909} is not present and does not play a role for the extended AFS.

\citet{Bruzek1969} stated that the length of an AFS is comparable with the width of a supergranule, which may indicate that an AFS covers a supergranule. In contrast, \citet{Zirin1974} claimed that an EFR and the associated AFS have no connections to the surrounding network. In the present case, the extended AFS connects to the flux at the border of a supergranule in the network, and it likely also covers the supergranule (Fig.~\ref{FIG:SolMon}c). This example shows that an AFS can connect magnetic flux from a sunspot with the flux of the network magnetic field, which potentially signifies the decay of the active region. Similarly, \citet{GonzalezManrique2017} described that an AFS connected a micro-pore with flux of the quiet-Sun, where the micro-pore evolved in a region of newly emerged flux.

The appearance of the filament in H$\alpha$, H$\beta$, and in AIA images is slightly different with respect to their morphology (Fig.~\ref{FIG:footpoints_overview}). In H$\alpha$ the opacity is higher than in H$\beta$. The structures appear very dark and single threads are distinguishable. In H$\beta$, on the other hand, the filament appears as a cloudy structure, and it is difficult to distinguish single threads. The photospheric pattern in the background is likely caused by the broad prefilter. After about 20\,minutes, the single arch filaments of the AFS are clearly separated. Nonetheless, H$\beta$ is a good alternative in the blue range of the spectrum for observations of filamentary structures. Furthermore, the arch filaments are well visible in the AIA images, although at lower spatial resolution. The clumpy absorption feature, which is visible in the strong chromospheric absorption lines, is missing in the central part of the AIA images (Fig.~\ref{FIG:footpoints_overview}c). We conclude that this filamentary structure, which connects the northern negative polarity (18\arcsec, 42\arcsec) and the southern positive polarity (18\arcsec, 34\arcsec), is located lower in the chromosphere compared to the rest of the AFS.

The analysis of LOS velocities is affected by the central wavelength at rest $\lambda_0$ chosen for calibration \citep[e.g.,][]{Kuckein2012b}. We derived LOS velocities for the chromospheric lines H$\alpha$ and NIR~\mbox{Ca\,\textsc{ii}}. The extended AFS was very dynamic in this short observing period. The maximum upflows in H$\alpha$ decreased in 30\,minutes from ($5.3\pm0.9$)\,km\,s$^{-1}$ to ($3.9\pm1.1$)\,km\,s$^{-1}$, whereas the maximum downflows increased from ($9.8\pm0.6$)\,km\,s$^{-1}$ to ($22.8\pm0.6$)\,km\,s$^{-1}$. The increase of redshifted profiles appeared at the location of the footpoints. In addition, the overall mean velocity evolved into an overall downflow of 0.7\,km\,s$^{-1}$ with a standard deviation of 2.6\,km\,s$^{-1}$ for the entire FOV. For NIR~\mbox{Ca\,\textsc{ii}} both blue- and redshifts increased in a time period of 30\,minutes. The upflows increased from ($4.3\pm0.3$)\,km\,s$^{-1}$ to ($7.7\pm0.3$)\,km\,s$^{-1}$ and the downflows from ($12.6\pm0.2$)\,km\,s$^{-1}$ to ($17.1\pm0.3$)\,km\,s$^{-1}$. At the end of the time-series, the up- and downflow velocities exceed the sound speed in the chromosphere \citep[e.g., Fig~2.5 in][]{LoehnerBoettcher2016phd}. These velocities were derived from line-core fits of the spectral line. Moreover, some H$\alpha$ and NIR~\mbox{Ca\,\textsc{ii}} spectral profiles show a second component, which is even more redshifted and leads to downward-directed velocities of about twice the velocities derived from the line core. A more detailed analysis of the second component is not possible because of the sparse wavelength sampling and the limited wavelength range.

By examining the spectra in Fig.~\ref{FIG:ibis_spectra}, we notice that the second component in the NIR~\mbox{Ca\,\textsc{ii}} spectra appears about five minutes later than in the H$\alpha$ profiles.  Furthermore, the area, in which a second-component appears and therefore supersonic velocities are reached, is much larger in H$\alpha$ than in NIR~\mbox{Ca\,\textsc{ii}} (Figs.~\ref{FIG:ibis_phys_6563} and \ref{FIG:ibis_phys_8542}). This indicates that the fast downstreaming plasma does not completely reach down to the lower chromospheric layer of NIR~\mbox{Ca\,\textsc{ii}}.

Previous observations of AFS in H$\alpha$ showed that loop tops are rising with velocities of 5\,--\,20\,km\,s$^{-1}$, whereas footpoint velocities of up to 20\,--\,50\,km\,s$^{-1}$ are measured \citep{Bruzek1969, Tsiropoula1992}. These values are in good agreement with the H$\alpha$ values of the extended AFS. Examining other chromospheric lines, \citet{GonzalezManrique2018} reported supersonic downflows at the footpoints of a single arch filament in \mbox{He\,\textsc{i}}\,10830\,\AA\ of up to 20\,--\,40\,km\,s$^{-1}$, whereby the average velocity decreased from 24\,km\,s$^{-1}$ to 1\,--\,7\,km\,s$^{-1}$ in only 7\,minutes. In our observations, we just see the increasing phase. Unfortunately, the observations were too short to cover the entire decreasing phase. Other studies examining  \mbox{He\,\textsc{i}}~10830\,\AA\ also identify supersonic velocities at the footpoints of up to 40\,\,km\,s$^{-1}$ \citep{Lagg2007, Balthasar2016}.

We find at the beginning of the time-series predominantly upstreaming flows in the central part of the clumpy AFS. The separated elongated filaments contain both  up- and downstreaming flows in the center and at the footpoints, respectively. Toward the end of the time-series, the downstreaming flows increased at the southern footpoint  (Fig.~\ref{FIG:ibis_phys_6563}). A similar behavior was already described by \citet{Frazier1972} in H$\alpha$ observations of an AFS, who described a rising arch filament, where the plasma is flowing down at its footpoints. \citet{GonzalezManrique2018} characterize the evolution of an individual arch filament. They observed in the \mbox{He\,\textsc{i}}~10830\,\AA\ line that the upflows at loop tops of the arch filament disappear and are replaced by downstreaming flows. They interpret the upflows as the rise of the whole arch filament until it is no longer detectable in \mbox{He\,\textsc{i}}~10830\,\AA, that is, when it enters the corona \citep[Fig.~15 in][]{GonzalezManrique2018}. Furthermore, they report an increase of the downstreaming flows at the footpoints of the arch filament. The same behavior is observed in H$\alpha$ and NIR~\mbox{Ca\,\textsc{ii}} data of our extended AFS, whereby the interpretation of \citet{GonzalezManrique2018} agrees well with our observations. 

We conjecture the following scenario for the development of the extended AFS in the vicinity of active region NOAA 11658: When the sunspot rotated into view on 2013~January~13, a large active region filament spread along the PIL between the positive polarity of the sunspot and the negative polarity of the surrounding network. As the active region filament decayed, an almost field-free region remained in which no arcade fields developed. This gave space so that loops of an arch filament system crossing the PIL could develop in this area. In the following days a strong moat flow developed around the sunspot (Fig.~\ref{FIG:SolMon}), which is inside the moat cell \citep{Strecker2018a}. The unpipolar MMFs in the moat flow transport flux away from the sunspot and are indicators of the decay of the sunspot. The MMFs moved toward the moat cell boundary and eventually merged with the magnetic network. Loops between the MMFs and the network flux along the boundary of a supergranule were deflected and developed a vertical component. In the following, these loops rose in the atmosphere as arch filaments. When the loop rises it carries plasma into higher chromospheric layers. Magnetic tension provides the restoring force for gravity. At some point, the magnetic field can no longer drag the plasma of the arch filament with it, and it drains toward the footpoint while, the loop continues its ascend. We speculate that this repeats over and over again on a time-scale of about 30\,minutes.

In AIA time-lapse movies (online movie referring to  Fig.~\ref{FIG:ibis_mag_align}), a brightening in the newly formed arch filament appears at around 17:20~UT. This brightening moves from the northern footpoint toward the southern one. We detect strong redshifts in the H$\alpha$ and NIR~$\mbox{Ca\,\textsc{ii}}$ filtergrams that appear after the brightening in AIA image at the southern footpoint. These brightenings are likely related to an activation of the AFS, which was too weak for the AFS to erupt. The cause for this activation was not detectable with the time-lapse movies of the entire active region. Nontheless, it was strong enough so that the spectra of H$\alpha$ and NIR~$\mbox{Ca\,\textsc{ii}}$ developed a second component with high redshifts.

Examining the temporal evolution of the downflows at the footpoints, we recognize an asymmetrical development. At 17:22\,UT, the maximum downstreaming velocity in H$\alpha$ at both footpoints is about the same with ($9.3\pm 0.5$)\,km\,s$^{-1}$ and ($9.8\pm1.5$)\,km\,s$^{-1}$ for the northern and southern footpoint, respectively. At the end of the time-series, the maximum downstreaming velocity at the northern footpoint decreased to ($5.9\pm0.4$)\,km\,s$^{-1}$, whereas  maximum downstreaming velocities of up to ($22.8\pm1.8$)\,km\,s$^{-1}$ were inferred at the southern footpoint. This demonstrates a strong asymmetry between both footpoints. This behavior was earlier reported for young AFSs with asymmetric vertical plasma flows of the leading and trailing legs of the arch filaments \citep{Spadaro2004}. Simulations of emerging flux tubes suggest an asymmetry \citep{Caligari1995}. \citet{Cauzzi1996} explain this asymmetry by the Coriolis force acting on the rising flux tube. In the aforementioned scenarios, the asymmetry was strongest at the beginning of the AFS's lifetime and decreased afterward. In the present extended AFS, the asymmetry is increasing over a short time period of 30\,minutes. This is related to the decaying active region itself and enhanced by the brightenings seen in the AIA images.

In the time-lapse movies (online movie referring to Fig.~\ref{FIG:ibis_mag_align}) of the IBIS Stokes-$V$ maps, we notice in the southern negative polarity part as well as in the northern positive polarity part flux cancellation. In this study, the LOS magnetic field in the AFS reaches values of more than 200\,G, where the negative polarity reaches even values of up to 600\,G based on photospheric HMI magnetograms. Exact values for the chromosphere could in principle be derived from inversions of the Stokes profiles. However, the signal-to-noise ratio of individual NIR~\mbox{Ca\,\textsc{ii}} Stokes profiles is too low for reliable spectral-line inversions.

In addition, we examined the rate at which the LOS magnetic flux changes based on the HMI magnetograms at the southern footpoint in the time period between 14\,--20\,UT. The flux decreased at a rate of $2.8 \times 10^{18}$\,Mx\,h$^{-1}$. The maximum flux at around 14\,UT is about $1.5 \times 10^{20}$\,Mx. In a typical emerging flux region, an AFS appears once the emerging flux is greater than $10^{20}$\,Mx and continues to increase \citep{Chou1988}. In the extended AFS, the flux decreases at the southern footpoint over time. Along with the MMFs in the moat flow around the penumbra, this indicates a decay of the sunspot.

When comparing the photospheric G-band bright points with the magnetic field in the photosphere (Fig.~\ref{FIG:footpoints_overview}), we find that the location of the G-band bright points is mainly close to the northern footpoint region, with some single bright points close to the southern footpoint. In the study of \citet{Ishikawa2007}, they discovered that G-band bright points are located at the border of strong small-scale flux elements, which is also visible in our observations. Furthermore, these authors point out that the strong small-scale flux elements surrounded by bright points show downflows, which is true for the present data set as well. We conclude that the bright points are related to strong magnetic flux but not necessarily to all footpoints of the extended AFS. \citet{Chae2005} argue that a careful study of the history of the magnetic field is important for the understanding of the relationship between the photospheric magnetic field region and endpoints of the filament.


\section{Conclusions}

We used DST data to examine the dynamics and evolution of an unusal AFS. The imaging spectroscopic data sets of IBIS provided information of the spectral lines in H$\alpha$ and NIR~\mbox{Ca\,\textsc{ii}} line. Here, we analyzed the temporal evolution with time-lapse movies of the extended AFS and its dynamics by an in-depth investigation of LOS velocities.

The extended AFS was atypical as compared to a classical AFS, which forms typically between emerging pores. The MMFs in the moat flow around the sunspot connect with opposite polarity flux of the surrounding network outside of the active region. This allows that the magnetic flux of the sunspot, which is already transported away by the MMFs, is carried to the magnetic network located further away. We refer to this filament system as an extended AFS because it evolves in an environment, where flux emergence has ceased long ago. Indeed, sunspot decay has already started. Time-lapse movies revealed that the compactly packed threads, by which the AFS appeared clumpy, separated over time and formed elongated arch filaments. Furthermore, a new arch filament formed during the one hour of observations. A clear polarity inversion line was absent in the Stokes-$V$ maps of the NIR~\mbox{Ca\,\textsc{ii}} line, which made it difficult to classify the chromopheric magnetic field structure. 

We determined the LOS velocities in the extended AFS. Here, we found upflows at the loop tops of an individual arch filament. At the footpoints, which were surrounded by G-band bright points, we detected downflows. Gradually, the upflows disappeared and were replaced by downflows. We interpret this as a rising flux tube. In H$\alpha$ the maximum upflows decreased from about 5\,km\,s$^{-1}$ to 3\,km\,s$^{-1}$ and the maximum downflows increased from about 10\,km\,s$^{-1}$ to supersonic values of about 23\,km\,s$^{-1}$. The downstreaming flows had asymmetrical velocities at both footpoints, which is not unusual for an AFS, but the asymmetry was intensified when a brightening of unknown origin became visible in the AIA images.

In the end, it was challenging to classify and analyze this extended arch filament system. High-resolution observations with high-temporal and spatial cadence and  polarization measurements with high sensitivity are necessary for further studies of this kind of AFS. With the next generation of solar telescopes, that is, Daniel K. Inouye Solar Telescope \citep[DKIST, ][]{Tritschler2015} and European Solar Telescope \citep[EST, ][]{Matthews2016}, new possibilities to study extended AFSs will arise. This will give more insight in the interaction of active regions with the surrounding magnetic network and the decaying mechanisms of sunspots.


\begin{acknowledgements}
The Dunn Solar Telescope is operated by the National Solar Observatory. The National Solar Observatory is operated by the Association of Universities for Research in Astronomy (AURA) under a cooperative agreement with the National Science Foundation. This publication makes use of data obtained during Cycle 1 of the DST Service Mode Operations under the proposal ID P100 (PI C. Kuckein). The authors thank Dr. Alexandra Tritschler for her support in designing and carrying out this DST service mode campaign. The Solar Dynamics Observatory was developed and launched by NASA. The SDO data are provided by the Joint Science Operations Center -- Science Data Processing at Stanford University. This research has made use of NASA's Astrophysics Data System. SolarSoftWare is a public-domain software package for analysis of solar data written in the Interactive Data Language by Harris Geospatial Solutions. This study is supported by the EU Horizon 2020 Research and Innovation Program under grant agreement number 824135 (SOLARNET). CD and CK acknowledge support by grant DE 787/5-1 of the German Research Foundation (DFG). The authors were partially supported by the German Academic Exchange Service (DAAD), with funds from the German Federal Ministry of Education \& Research and Slovak Academy of Science, under project No. 57449420.  The authors thank Drs. S. J. Gonz\'alez Manrique, I. Kontogiannis, and H. Balthasar for their thoughtful reading of the manuscript and the helpful comments. The authors thank the referee for constructive criticism and helpful suggestions improving the manuscript.
\end{acknowledgements}

\bibliographystyle{aa}
\bibliography{apj-jour.bib,Diercke_eAFS.bib}

\end{document}